\documentclass[journal]{IEEEtran}
\usepackage{amsmath,amssymb,amsfonts}
\usepackage{algorithm}
\usepackage{algorithmicx}
\usepackage{algpseudocode} 
\usepackage{array}
\usepackage[caption=false,font=footnotesize,labelfont=rm,textfont=rm]{subfig}
\usepackage{textcomp}
\usepackage{stfloats}
\usepackage{url}
\usepackage{verbatim}
\usepackage{graphicx}
\usepackage{cite}
\usepackage{balance}
\usepackage{makecell}
\usepackage{flushend}
\begin{document}
\flushend
\title{A Hybrid Labeled Multi-Bernoulli Filter With Amplitude For Tracking Fluctuating Targets}
\author{\IEEEauthorblockN{Weizhen Ma, Zhongliang Jing, Peng Dong, Henry Leung}
\thanks{This work is supported by National Natural Science Foundation of China under Grants 61803260 and 61673262. \emph{(Corresponding authors: Zhongliang Jing; Peng Dong.)}} 
\thanks{Weizhen Ma, Zhongliang Jing and Peng Dong are with the School of Aeronautics and Astronautics, Shanghai Jiao Tong University, Shanghai 200240, China (e-mail: weizhenma@sjtu.edu.cn; zljing@sjtu.edu.cn; dongpengkty@sjtu.edu.cn).}
\thanks{Henry Leung is with the Department of Electrical and Computer Engineering, University of Calgary, Calgary AB T2N 1N4, Canada (e-mail: leungh@ucalgary.ca).}}
\maketitle

\markboth{IEEE Transactions On Signal Processing } {Ma \MakeLowercase{\textit{et al.}}: A Hybrid Labeled Multi-Bernoulli Filter With Amplitude For Tracking Fluctuating Target}


\begin{abstract}
The amplitude information of target returns has been incorporated into many tracking algorithms for performance improvements. One of the limitations of employing amplitude feature is that the signal-to-noise ratio (SNR) of the target, i.e., the parameter of amplitude likelihood, is usually assumed to be known and constant. In practice, the target SNR is always unknown, and is dependent on aspect angle hence it will fluctuate. In this paper we propose a hybrid labeled multi-Bernoulli (LMB) filter that introduces the signal amplitude into the LMB filter for tracking targets with unknown and fluctuating SNR. The fluctuation of target SNR is modeled by an autoregressive gamma process and amplitude likelihoods for Swerling 1 and 3 targets are considered. Under Rao-Blackwell decomposition, an approximate Gamma estimator based on Laplace transform and Markov Chain Monte Carlo method is proposed to estimate the target SNR, and the kinematic state is estimated by a Gaussian mixture filter conditioned on the target SNR. The performance of the proposed hybrid filter is analyzed via a tracking scenario including three crossing targets. Simulation results verify the efficacy of the proposed SNR estimator and quantify the benefits of incorporating amplitude information for multi-target tracking.       

\end{abstract}

\begin{IEEEkeywords}
  Multi-target tracking, LMB filter, amplitude information, SNR fluctuation, Swerling target
\end{IEEEkeywords}

\section{Introduction}
\indent \IEEEPARstart{I}{n} multi-target tracking, conventional measurements of range, azimuth, Doppler, etc., are usually used to establish and maintain target tracks. Due to the observation noise of the sensor and false alarms, the received measurements are always imprecise and contain background clutters, which poses great challenges to tracking applications. However, sensors such as radar and sonar that typically provide the position measurement can also output the strength or amplitude of the target signal \cite{LB1990}. Since the amplitudes of target returns are usually stronger than those of false alarms, amplitude information has the ability to discard false alarms, improve measurement to track association and tracking performance. Target amplitude has been employed for tracking target in cluttered environment \cite{LB1990,LB1993,V1996,CRVV2010,BHG2010}, closely spaced targets \cite{YWLL2012, MUK2016}, and track-before-detect algorithms \cite{TB1998,KRGR2021,RKRG2022}. \\
\indent In the pioneering works of Lerro and Bar-Shalom, the probabilistic data association filter with amplitude information (PDA-AI) is developed for single target tracking in cluttered environment \cite{LB1990}. Then they combine the PDA-AI with interacting multiple model for tracking maneuvering target \cite{LB1993}. Subsequently, Van Keuk extends signal strength to multiple hypothesis tracking and analyzes the influence of target signal-to-noise ratio (SNR) on the selection of detection threshold and tracking performance \cite{V1996}. \\
\indent The implementations of the aforementioned methods rely on the prior knowledge of the target SNR, which is usually unknown and has to be estimated in practice. In \cite{CRVV2010}, a simplified amplitude likelihood for unknown target SNR is proposed by marginalizing the conditional likelihood over a possible SNR interval. The marginalized likelihood is independent of the target SNR and is introduced into the probability hypothesis density (PHD) and cardinalized PHD (CPHD) filters. A more robust PDA-AI filter is proposed in \cite{BHG2010} for tracking small maritime target in heavy-tailed backgrounds. The authors replace the conventional Rayleigh amplitude likelihood used in \cite{LB1990,LB1993,V1996,CRVV2010} with $K$-distribution for sea clutter, and suggest estimating the SNRs of the background noise and the target by moment-based method. The $K$-distributed amplitude likelihood is then combined with the labeled multi-Bernoulli (LMB) filter in \cite{SLLC2019} and the Bernoulli track-before-detect filter in \cite{KRGR2021} both for marine target tracking. Moreover, signal amplitude has also been exploited to estimate the average radar cross-section (RCS), which is used as an additional attribute to enhance the CPHD filter for tracking closely spaced targets \cite{MUK2016}.\\ 
\indent The target SNR is in fact proportional to the mean RCS of a target \cite{MUK2016}, hence they can be treated equivalently. The fluctuation of the target RCS is typically described by the Swerling models \cite{S1960} with the average RCS being the model parameter. However, some literature has revealed that radar targets exhibit different statistical properties with respect to aspect angle, and suggested that the average RCS should be specified for different aspect angle rather than being fixed \cite{CRVV2010,W1972,J1997,BKM2014,M2019}. Hence, one has to take into account the fluctuation of the target SNR when amplitude information is used. In \cite{BKLYS2010,B2019}, the uncertainty of the target SNR is modeled by random walk (RW) with Gaussian step size and the SNR is estimated by sequential Monte Carlo (SMC) method. Ristic et al. \cite{RRKG2021} employ the Bernoulli filter with amplitude measurement for single target tracking, where the target SNR is also described by a RW process. Using the Rao-Blackwellisation principle \cite{MR2000}, the joint state density is decomposed into the densities of the target SNR and the kinematic state conditioned on SNR, then they are estimated by particle filter and Gaussian mixture (GM) filter, respectively. However, the estimation of the particle dependent spatial state suffers from computational burden as the number of particles grows. In \cite{M2019}, a noncentred Gamma (NCG) distribution \cite{GJ2006} is used to describe the evolution of the hidden state, which is inversely proportional to the local average RCS and induces RCS measurements. Under the assumptions that the observed RCS and the prior state density both follow a Gamma distribution, a closed-formed tracker is proposed to estimate the time varying local average RCS. The NCG distribution provides an alternative process model for the fluctuation of the target SNR.\\ 
\indent In this paper, we aim to track targets with unknown and fluctuating SNR using the LMB filter with amplitude information. The resulting hybrid LMB (HLMB) filter is formulated according to the Rao-Blackwell decomposition: the target SNR is estimated via a Gamma filter while the conditional kinematic state is estimated via a GM filter. The HLMB filter with Gamma approximation of the target SNR has the potential for computational load reduction compared with the conventional SMC implementation of the SNR. \\
\indent The Gamma estimator is constructed under the assumptions that the SNR transition density follows a NCG distribution and the prior SNR density follows a Gamma distribution. The predictive SNR density is approximated by Gamma probability density function (PDF) using Laplace transform and moment matching. For either Rayleigh (Swerling 1 target) or one-dominant-plus-Rayleigh (Swerling 3 target) amplitude likelihood \cite{DR1980}, the Bayesian posterior density of the target SNR has a rather complicated expression and includes an intractable normalizing constant. To mitigate these limitations, the Markov Chain Monte Carlo (MCMC) method is used to approximate posterior SNR density with Gamma distribution. Then the estimation of the kinematic state can be carried out conditioned on the minimum mean square error (MMSE) estimate of the target SNR. Simulation results of the proposed hybrid filter are presented to demonstrate the efficacy of the SNR estimator and the improved tracking performance.  \\
\indent This paper is organized as follows. Section \ref{sec_two} introduces some basic notations and the models for SNR fluctuation and amplitude likelihood. Section \ref{sec_three} demonstrates the proposed HLMB filter. Under Rao-Blackwell decomposition, the conditional kinematic state is estimated by a GM filter. In Section \ref{sec_four}, we first give a particle representation of the SNR density, then propose an Gamma estimator of the target SNR based on Laplace transform and MCMC method. Simulation results are presented in Section \ref{sec_five} and the conclusions are summarized in Section \ref{sec_six}.    
\section{Notation and Model} \label{sec_two} 
\subsection{Notation}
\indent In this paper, single-target states are represented by lowercase letters (e.g., $x$, $\mathbf{x}$) and multi-target states are given by uppercase letters (e.g.,$X$, $\mathbf{X}$). The bolded symbols are used for labeled states and distributions, e.g., $\mathbf{x}$, $\mathbf{X}$, $\boldsymbol{\pi}$, to distinguish from their unlabeled counterparts. Moreover, let blackboard bold letters denote spaces, e.g., $\mathbb{X}$ denotes state space and $\mathbb{Z}$ denotes measurement space, and let $\mathcal{F}(\mathbb{S})$ represent the collection of all subsets of $\mathbb{S}$.\\
\indent The following abbreviations are used in the paper for notational convenience. The inner product of two continuous functions $f(x)$ and $g(x)$ is denoted by $\langle f,g\rangle \triangleq\int f(x)g(x)\mathrm{d}x$. The multi-target exponential notation is given by
\begin{equation}
    h^X\triangleq\prod_{x\in X}h(x),
\end{equation}
where $h$ is a real-valued function and $h^X =1$ in case of $X=\varnothing$. Besides, the generalized Kronecker delta function
\begin{equation}
    \delta_Y(X) \triangleq  
    \begin{cases} 
        1,&\text{if}~X=Y\\ 
        0,&\text{otherwise} 
    \end{cases}
\end{equation}
facilitates the application of Kronecker delta function to integers, vectors and sets. The inclusion function
\begin{equation}
    1_Y(X)\triangleq
    \begin{cases}
        1,~\text{if}~X\subseteq Y\\
        0,~\text{otherwise}
    \end{cases}
\end{equation}
indicates whether a set $X$ is a subset of $Y$.  \\
\indent The LMB filter \cite{SBBK2014,DJLSL2019,SDJL2021,SZDJL2022} used in this paper is a suboptimal Bayesian multi-target tracking algorithm, which characterizes the multi-target density by an LMB random finite set (RFS) $\mathbf{X}=\{(r^\ell,p^\ell(x))\}_{\ell\in\mathbb{L}}$, where $r^\ell$ and $p^\ell(x)$ are the existence probability and spatial density for the target with label $\ell$, which belongs to the label space $\mathbb{L}$. The density of $\mathbf{X}$ is given by
\begin{equation}
    \boldsymbol{\pi}({\mathbf{X}})=\Delta(\mathbf{X})w(\mathcal{L}(\mathbf{X}))p^\mathbf{X},
\end{equation}
where $\Delta(\mathbf{X})=\delta_{|\mathbf{X}|}(\mathcal{L(\mathbf{X})})$, $\mathcal{L}(\mathbf{X})$ is the label set of $\mathbf{X}$, $\mathcal{L}$ is the projection defined by $\mathcal{L}(\mathbf{x})=\mathcal{L}((x,\ell))=\ell$ and 
\begin{equation}\label{eqn_LMB_weight}
    w(L)=\prod_{i\in\mathbb{L}}(1-r^{i})\prod_{\ell\in L}\frac{1_{\mathbb{L}}(\ell)r^{\ell}}{1-r^{\ell}},
\end{equation}
\begin{equation}
    p(x,\ell)=p^{\ell}(x).
\end{equation}
\subsection{SNR Fluctuating Model}
\indent We consider an augmented target state $x=[p_1~ \dot{p}_1~p_2~\dot{p}_2~d]^T$, which consists of the target positions $p_1$ and $p_2$, their velocities $\dot{p}_1$ and $\dot{p}_2$, and the target SNR $d$. The LMB filter also assigns each target a unique and invariant track label $\ell$, leading to the labeled state $\mathbf{x}=(x,\ell)$. The state set at time $k$ is $X_k=\{x_{k}^1,...,x_{k}^{|X_k|}\}$, where $|X_k|$ is the number of targets. Since the kinematic state and target SNR will be treated separately, the notation $\tilde{x}=[p_1~ \dot{p}_1~p_2~\dot{p}_2]^T$ is used for the kinematic state, then we have $x=[\tilde{x}^T~d]^T$. \\
\indent In practice, the target SNR is dependent on the aspect angle and radar wave frequency \cite{W1972,BKM2014}, hence it will fluctuate. The dynamics of SNR has been simply described by random walk model or Gaussian density \cite{BKLYS2010,RRKG2021}. Alternatively, we employ the NCG distribution as the SNR transition density, which is well-known for modeling stochastic volatility such as financial time series \cite{GJ2006} and wind speed intensity \cite{CP2012}. Then the evolution of SNR follows 
\begin{equation}\label{eqn_ncgamma}
    \begin{aligned}
        &f_d(d_k | d_{k-1};\delta,\rho,c)=\\
        &\begin{cases}
            0,&d_k\leq 0\\
            \exp \left( -\frac{d_k}{c} \right) \sum_{i=0}^{\infty}{\frac{d_{k}^{\delta +i-1}}{c^{\delta +i}\Gamma (\delta +i)}}\frac{e^{-\rho d_{k-1}/c}(\rho d_{k-1}/c)^i}{i!},&d_k> 0\\
        \end{cases} 
    \end{aligned}
\end{equation}
which is characterized by the degree of freedom $\delta$, autoregressive coefficient $\rho$, scale parameter $c$ and Gamma function $\Gamma(\cdot)$. The resulting sequence $\{d_k\}$ is an autoregressive Gamma (ARG) process with stationary distribution, for $\rho<1$, being the Gamma distribution $\gamma_s(\delta,c/(1-\rho))$ given by
\begin{equation}
    \gamma_s \left( x;\varphi,\theta \right) =
    \begin{cases}
        0,&x\leq 0\\
        \frac{1}{\Gamma (\varphi)\theta^\varphi}x^{\varphi-1}e^{- x/\theta},& x>0
    \end{cases}
\end{equation}
where $\varphi$ is the shape parameter and $\theta$ is the scale parameter. The NCG distribution can be interpreted as a Poisson mixture of Gamma distributions $\gamma_s(\delta+i,c)$ with probability weights $p_i=\lambda^ie^{-\lambda}/i!$ for $i=1,2,...,$ where $\lambda = \rho d_{k-1}/c$. \\
\indent Although the transition model \eqref{eqn_ncgamma} has a rather complicated expression, the equivalent ARG process can be easily simulated by sampling from Gamma and Poisson random variables. In particular, the trajectory of $d_k$ is generated by \cite{S1990}
\begin{equation}\label{eqn_AR1}
    d_k=\sum_{i=1}^{N(d_{k-1})}{W_{i,k}}+\varepsilon _k, 
\end{equation}
where $W_{i,k}$ are Gamma random variables sampled from $\gamma_s(1,c)$, $N(d_{k-1})$ is extracted from a Poisson density with rate parameter $\rho d_{k-1}/c$, the residual $\varepsilon _k$ follows a Gamma density $\gamma_s(\delta,c)$. Moreover, the conditional mean and variance of the NCG distribution are
\begin{equation}\nonumber
    E(d_k|d_{k-1})=c\delta+\rho d_{k-1},~ V(d_k|d_{k-1})=c^2\delta+2\rho cd_{k-1}.
\end{equation}
The ARG process, with three adjustable parameters, offers more flexibility to simulate the behavior of SNR compared to the Gaussian density characterized by mean and standard deviation. It can be underdispersed or overdispersed by choosing different values of parameters \cite{GJ2006}. \\
\indent Under the assumption that $\delta\rightarrow 0$ and $\rho=1$ in \cite{M2019}, the martingale property of the process $\{d_k\}$ can be obtained, i.e., $E(d_k|d_{k-1})=d_{k-1}$. The conditional variance also simplified to $V(d_k|d_{k-1})=2cd_{k-1}$, which suggests that the target SNR will drift slowly for small values of $c$. However, this assumption is not used here since the residual term $\varepsilon_k$ will be excluded from $\eqref{eqn_AR1}$ as $\delta\rightarrow 0$, and the stationarity of the ARG process is not satisfied as $\rho=1$. Instead, we assume $\delta=1$ for Swerling 1 target and $\delta=2$ for Swerling 3 target, and set $\rho\rightarrow 1$. Then an approximate martingale ARG process can be obtained for small values of $c$. \\
\indent In most tracking applications, it is reasonable to assume that the dynamics of the kinematic state $\tilde{x}$ is independent of the target SNR $d$ \cite{RRKG2021}. Then the transition density of the augmented state $x$ can be decomposed as follows
\begin{equation}\label{eqn_state_trans}
    f(x|x')=f(\tilde{x},d|\tilde{x}',d')=f_{\tilde{x}}(\tilde{x}|\tilde{x}')f_d(d|d'),
\end{equation}
with $f_{\tilde{x}}(\tilde{x}|\tilde{x}')=\mathcal{N}(\tilde{x};F\tilde{x}',Q)$, where $\mathcal{N}(x;\tilde{x},\Sigma)$ denotes the Gaussian PDF with mean $\tilde{x}$ and covariance matrix $\Sigma$, matrices $F$ and $Q$ represent the transition matrix and process noise covariance matrix, respectively.
\subsection{Amplitude Likelihood}
\begin{figure}
    \centering
    \includegraphics[width=3in]{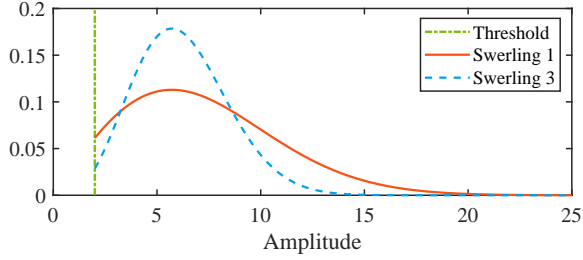}
    \caption{Amplitude likelihoods for target with $\text{SNR} = 15$ dB and $\tau=2$ for Swerling 1 and Swerling 3.}
    \label{fig_amplitude_lh}
\end{figure}

\indent The measurement set at time $k$ is $Z_k=\{z_{k}^1,...,z_{k}^{|Z_{k}|}\}$, and the measurement sequence up to time $k$ is denoted by $Z_{1:k}=(Z_1,...,Z_k)$. Each measurement $z=[\tilde{z}^T~a]^T$ consists of the two dimensional Cartesian position vector $\tilde{z}$ induced by target states $p_1$ and $p_2$,  and the signal amplitude $a$ induced by the target SNR $d$. Throughout this paper, the amplitude refers to the output of a bandpass matched filter followed by an envelope detector \cite{DR1980,LB1990}. The target amplitude is typically assumed to fluctuate from scan-to-scan or pulse-to-pulse \cite{S1960}. We consider the scan-to-scan condition such that the amplitude is constant during a single scan but fluctuates independently from scan to scan. For the Swerling 1 target, which consists of many individual scatterers without a dominant one, its amplitude follows a Rayleigh PDF \cite{DR1980}
\begin{equation}\label{eqn_amplsone}
    p^1\left( a\mid d\right) =\frac{a}{1+d}\exp\left(-\frac{a^2}{ 2\left(1+d\right)}\right), 
\end{equation}
where $d$ is the target SNR, which is typically defined in the logarithmic scale as $\text{SNR(dB)}=10\log_{10}(1+d)$. For the target consisting of many small scatterers with a dominant scatterer, i.e., the Swerling 3 target, the PDF of its amplitude is the one-dominant-plus-Rayleigh distribution \cite{DR1980}
\begin{equation}\label{eqn_amplsthree}
    p^3\left( a\mid d\right) =\frac{9a^3}{2\left( 1+d \right) ^2}\exp\left(-\frac{3a^2}{2\left(1+d\right)}\right) .
\end{equation}
\indent The detection probability of a target can be defined as the probability that its signal strength exceeds a given detection threshold $\tau$, i.e.,
\begin{equation}\label{eqn_PD}
    P_{D}^{\kappa}\left( d,\tau \right) =p^\kappa\left( a>\tau\mid d\right) =\int_{\tau}^{\infty}{p^\kappa}\left( a\mid d\right) \mathrm{d}a, 
\end{equation}
where $\kappa=1,3$ indicates the type of Swerling model. Inserting \eqref{eqn_amplsone} and \eqref{eqn_amplsthree} into \eqref{eqn_PD}, we have
\begin{equation}
    \begin{aligned}
        P_{D}^{1}\left(d, \tau\right)&=\exp\left(-\frac{\tau^2}{2\left(1+d\right)} \right),\\
        P_{D}^{3}\left(d, \tau\right)&=\left(1+\frac{3\tau^2}{2\left(1+d\right)^{2}}\right)\exp\left(-\frac{3\tau^2}{2\left(1+d\right)} \right).
    \end{aligned}
\end{equation}
\indent Due to the introduction of detection threshold $\tau$, the amplitude likelihoods given by \eqref{eqn_amplsone} and \eqref{eqn_amplsthree} has to be renormalized for $a>\tau$, i.e.,
\begin{equation}\label{eqn_amplitude_lh}
    \begin{aligned}
        p^\kappa_{\tau}\left(a \mid d\right) &=p^\kappa\left(a \mid a>\tau, d\right) =\frac{p^\kappa\left(a \mid d\right)}{P_{D}^{\kappa}\left(d, \tau\right)},\\
        p^1_{\tau}\left(a \mid d\right)&=\frac{a}{1+d}\exp\left(\frac{\tau^2-a^2}{2\left(1+d\right)}\right),\\
        p^3_{\tau}\left(a \mid d\right)&=\frac{9a^3}{3\tau^2(1+d)+2(1+d)^2}\exp\left(\frac{3(\tau^2-a^2)}{2\left(1+d\right)} \right).
    \end{aligned}
\end{equation}
\indent Fig. \ref{fig_amplitude_lh} shows the amplitude likelihoods for a target with SNR $d=15$ dB and threshold $\tau=2$. It can be seen that the amplitude values of the Swerling 3 target are more concentrated closely near the mean than the values of Swerling 1 target. Hence the Swerling 3 amplitude likelihoods for targets with different SNR will have fewer overlapping regions and provide better discrimination between targets. Under the assumption that the background noise is normalized \cite{LB1990}, the amplitude densities for clutter, denoted as $c^\kappa_\tau(a)$, can be obtained by setting $d=0$ in \eqref{eqn_amplitude_lh}, i.e., $c^\kappa_\tau(a)=p^\kappa_{\tau}\left(a \mid 0\right)$. \\
\indent After incorporating amplitude information into the measurement vector, the multi-target likelihood is modified with the amplitude likelihood ratio $p^\kappa_{\tau}\left(a \mid d\right)/c^\kappa_\tau(a)$ \cite{CRVV2010}. The single target measurement likelihood can be decomposed as
\begin{equation}\label{eqn_meas_lh}
    g(z|x)=g(\tilde{z},a|\tilde{x},d)=g_{\tilde{z}}(\tilde{z}|\tilde{x})p^\kappa_{\tau}(a|d)
\end{equation}
where $g_{\tilde{z}}(\tilde{z}|\tilde{x})=\mathcal{N}(\tilde{z};H\tilde{x},R)$ is the likelihood of the position measurement, matrices $H$ and $R$ are the observation matrix and the observation noise covariance, respectively.
\section{The Hybrid Labeled Multi-Bernoulli Filter}\label{sec_three}
\indent The LMB filter is an efficient approximation of the generalized labeled multi-Bernoulli filter which is an optimal but complex multi-target tracking algorithm. In an attempt to incorporate amplitude information into the LMB filter for tracking fluctuating targets, we develop a hybrid LMB filter that simultaneously estimates the spatial state and the SNR of the target. The proposed hybrid filter takes advantage of the target amplitude and belongs to the conventional point-measurement Bayesian filtering framework. The density of the augmented target state is divided into two parts using Rao-Blackwell decomposition, such that the two densities can be treated separately for computational complexity reduction. 
\subsection{HLMB Recursion}
\indent The HLMB filter with target amplitude, reformulated on the augmented state and measurement spaces, is a straightforward extension of the LMB filter. Thus, the derivation of it can be carried out similarly to the LMB filter. Suppose the prior multi-target density and birth density are LMB RFSs with parameter sets $\boldsymbol{\pi}=\{(r^\ell,p^\ell(x))\}_{\ell\in\mathbb{L}}$ and $\boldsymbol{\pi}_B=\{(r^\ell_B,p^\ell_B(x))\}_{\ell\in\mathbb{B}}$, respectively, then the predictive density is still an LMB RFS given by
\begin{equation}
    \boldsymbol{\pi }_+=\{( r_{+}^\ell,p_{+}^\ell(x) ) \}_{\ell\in \mathbb{L}}\cup \{( r_{B}^\ell,p_{B}^\ell(x) ) \}_{\ell\in \mathbb{B}},
\end{equation}
where
\begin{equation} \label{eqn_r_predict}
    r_{+}^\ell=\eta _S\left( \ell \right) r^\ell,
\end{equation}
\begin{equation}
    \begin{aligned}
        p_{+}^\ell(x)&=\frac{\langle P_S\left( x' ,\ell \right) f\left( x|x'  \right) ,p^\ell\left( x' \right) \rangle}{\eta _{S}\left( \ell \right)}\\
        &=\frac{\iint P_S\left( \tilde{x}',d' ,\ell \right)f_{\tilde{x}}(\tilde{x}|\tilde{x}')f_d(d|d')p^\ell(\tilde{x}',d')\mathrm{d}\tilde{x}'\mathrm{d}d'}{\eta _{S}\left( \ell \right)},\\
    \end{aligned}
\end{equation}
\begin{equation}
    \eta _{S}\left( \ell \right) =\iint P_S\left( \tilde{x}',d' ,\ell \right) ,p^\ell\left( \tilde{x}',d' \right) \mathrm{d}\tilde{x}'\mathrm{d}d',\\
\end{equation}
and $P_S\left( x ,\ell \right)$ is the state dependent survival probability. The label space of the predictive density, i.e., $\mathbb{L}_+=\mathbb{B}\cup \mathbb{L}$, satisfies $\mathbb{B}\cap \mathbb{L}=\varnothing $. \\
\indent If the multi-target predictive density is an LMB RFS represented by parameter set $\boldsymbol{\pi}_+=\{(r_+^\ell,p_+^\ell)\}_{\ell\in\mathbb{L}_+}$, then the filtering density can be approximated by an LMB RFS, i.e., $\boldsymbol{\pi}(\mathbf{X}\vert Z)\approx\{(r^\ell,p^\ell(x))\}_{\ell\in\mathbb{L}_+}$, where
\begin{equation}\label{eqn_r_post}
    r^{\ell}=\sum_{\left( I_+,\theta \right) \in \mathcal{F}\left( \mathbb{L}_+ \right) \times \Theta _{I_+}}{w^{\left( I_+,\theta \right)}}\left( Z \right) 1_{I_+}\left( \ell \right),
\end{equation}
\begin{equation}\label{eqn_p_post}
    p^{\ell}\left( x \right) =\frac{1}{r^{\ell}}\sum_{\left( I_+,\theta \right) \in \mathcal{F}\left( \mathbb{L}_+ \right) \times \Theta _{I_+}}{w^{\left( I_+,\theta \right)}}\left( Z \right) 1_{I_+}\left( \ell \right) p^{\left( \theta \right)}\left( x,\ell \right),
\end{equation}
and $I_+=\{\ell_1,\dots,\ell_{|I_+|}\}$ denotes the label set of the predicted tracks, $\Theta_{I_+}$ is the space mappings $\theta:I_+\rightarrow\{0,1,\dots,|Z|\}$ with $0$ represents dummy measurement, $Z$ is the set of observations at current step where the time index is omitted unless needed for clarity in what follows, and
\begin{equation}
    {w^{\left( I_+,\theta \right)}}\left( Z \right)\propto w_+(I_+)\bigg[\eta_Z^{(\ell)}(x;\theta)\bigg]^{I_+},
\end{equation}
\begin{equation}\label{eqn_state_posterior}
    p^{\left( \theta \right)}\left( x,\ell \right) =\frac{p_+^\ell\left( x \right) \psi^\ell _Z\left( x;\theta \right)}{\eta _{Z}^{\left( \ell \right)}\left( x;\theta \right)},
\end{equation}
\begin{equation}
\eta _{Z}^{\left( \ell \right)}\left( x;\theta \right) =\left< p_+^\ell\left( \cdot  \right) ,\psi^\ell _Z\left( \cdot;\theta \right) \right>,
\end{equation}
\begin{equation}
    \psi^\ell_Z\left( x;\theta \right) =
    \begin{cases}
        \frac{P_{D}^{\kappa}\left(d, \tau\right) g\left( z^{\theta \left( \ell \right)}|x \right)}{\varphi \left( \tilde{z} \right)c^\kappa_\tau(a)}, &\theta \left( \ell \right) >0\\
        1-P_{D}^{\kappa}\left(d, \tau\right) ,&\theta \left( \ell \right) =0\\
    \end{cases}
\end{equation}
where $w_+(I_+)$ is the weight of label set $I_+$ given by \eqref{eqn_LMB_weight}, $P_{D}^\kappa(d,\tau)$, $g(z|x,\ell )$ and $c^\kappa_\tau(a)$ have been introduced in Section \ref{sec_two}, $\varphi(\cdot)$ is the intensity of Poisson clutter. The likelihood $\eta_{Z}^{\left( \ell \right)}(x;\theta)$ for assigning measurement $z^{\theta(\ell)}=[\tilde{z}^T~ a]^T$ to track $\ell$ forms the cost matrix $C$ of the HLMB filter \cite{BBD2014}, from which we can extract the most likely association hypotheses $(I_+,\theta)$ by their weights ${w^{\left( I_+,\theta \right)}}\left( Z \right)$ instead of taking all hypotheses into account in \eqref{eqn_r_post} and \eqref{eqn_p_post}. 
\subsection{Rao-Blackwell decomposition}
\indent The GM implementation provides an analytical solution to the LMB filter under the linear Gaussian assumptions of the target dynamics and measurement model. However, since the SNR fluctuating model and amplitude likelihood are both non-Gaussian, the solution for the HLMB filter has to be reformulated. Rao-Blackwellisation facilitates the decomposition of the joint posterior $p^{\ell}(x)=p^{\ell}(\tilde{x},d)$ when two conditions are satisfied \cite{MR2000,RRKG2021}. First, the transition density of augmented state $x$ can be decomposed as
\begin{equation}\label{eqn_rb_transition}
    \begin{aligned}
        f(\tilde{x},d|\tilde{x}',d')&=f_{\tilde{x}}(\tilde{x}|\tilde{x}',d,d')f_d(d|x',d')\\
        &=f_{\tilde{x}}(\tilde{x}|\tilde{x}',d,d')f_d(d|d').
    \end{aligned}
\end{equation}
The second equation in \eqref{eqn_rb_transition} follows the assumption that the fluctuation of the target SNR is independent of its kinematic state, see \eqref{eqn_ncgamma}. Bseides, $f_{\tilde{x}}(\tilde{x}|\tilde{x}',d,d')$ can be simplified to $f_{\tilde{x}}(\tilde{x}|\tilde{x}')$, see \eqref{eqn_state_trans}. Second, the conditional posterior density $p^\ell(\tilde{x}|d)$ is analytically tractable. This condition can be satisfied if the SNR $d$ is known and the approximate solution is the GM implementation of the LMB (GM-LMB) filter \cite{SBBK2014}. Then the kinematic state $\tilde{x}$ can be marginalized out from the posterior $p^{\ell}(\tilde{x},d)$ so that we only have to concentrate on solving $p^\ell(d)$. \\ 
\indent Formally, we make use of the following decomposition of the joint posterior
\begin{equation}\label{eqn_raoblackwell}
    p^{\ell}(x)=p^{\ell}(\tilde{x},d)=p^\ell(\tilde{x}|d)p^\ell(d).
\end{equation}
This factorization allows us to first deal with the target SNR, then compute the kinematic state based on the estimated SNR. Given a measurement to track association mapping $\theta$, the SNR density admits the following recursion
\begin{equation}\label{eqn_snr_predict}
    p^\ell_+(d)=\int f_d(d|d')p^\ell(d')\mathrm{d}d',
\end{equation}
\begin{equation}\label{eqn_snr_update}
    p^{(\theta)}(d,\ell)=\frac{p(z^{\theta(\ell)}|d)p^\ell_+(d)}{\int p(z^{\theta(\ell)}|d)p^\ell_+(d)\mathrm{d}d}=\frac{p^\kappa_{\tau}(a|d)p^\ell_+(d)}{\int p^\kappa_{\tau}(a|d)p^\ell_+(d)\mathrm{d}d}.
\end{equation}
The second equation in \eqref{eqn_snr_update} holds because the position measurement $\tilde{z}$ is independent of SNR $d$, see \eqref{eqn_meas_lh}. Given all association hypothesis $\{(I_+,\theta)\}$, the posterior SNR density $p^\ell(d)$ can be calculated by substituting $x$ with $d$ in \eqref{eqn_p_post}. The conditional posterior $p^\ell(\tilde{x}|d)$ will be solved analytically by a GM filter in the next subsection and the marginal posterior $p^{(\ell)}(d)$ will be solved approximately by a Gamma estimator in the next section.
\subsection{GM Implementation of the Kinematic State} \label{sec_gm}
\indent The conditional density $p^\ell(\tilde{x}|d)$ is analytically tractable under linear Gaussian models, here we provide its GM implementation. Suppose the conditional posterior is in the form of GM, i.e., $p^\ell(\tilde{x}|d)=\sum_{i=1}^{J(\ell)} w_{i}^{\ell,\tilde{x}} \mathcal{N}\left(\tilde{x} ; m_{i}^{\ell}, P_{i}^{\ell}\right)$, and the survival probability is a constant, i.e., $P_S(x,\ell)=P_S$, then we have
\begin{equation}
    \eta _{S}\left( \ell \right)=P_S\iint p^\ell(\tilde{x}'|d')p^\ell(d')\mathrm{d}\tilde{x}'\mathrm{d}d'=P_S,
\end{equation}
\begin{equation}
    \begin{aligned}
        p_{+}^\ell(x)&=\frac{P_S\int f_{\tilde{x}}(\tilde{x}|\tilde{x}')p^\ell(\tilde{x}'|d')\mathrm{d}\tilde{x}' \int f_d(d|d')p^\ell(d')\mathrm{d}d'}{P_S}\\
        &=\int f^\ell_{\tilde{x}}(\tilde{x}|\tilde{x}')p^\ell(\tilde{x}'|d')\mathrm{d}\tilde{x}' p_+^\ell(d)\\
        &=p^\ell_+(\tilde{x}|d')p^\ell_+(d).
    \end{aligned}
\end{equation}
The predictive spatial density is also in the form of GM 
\begin{equation} \label{eqn_x_predict}
    p^\ell_+(\tilde{x}|d')=\sum_{i=1}^{J(\ell)} w_{i,+}^{\ell,\tilde{x}} \mathcal{N}\left(\tilde{x} ; m_{+,i}^{\ell}, P_{+,i}^{\ell}\right),
\end{equation}
where $w_{i,+}^{\ell,\tilde{x}}=w_{i}^{\ell,\tilde{x}}$, $\quad m_{+,i}^{\ell}=Fm_{i}^{\ell}$ and $\quad P_{+,i}^{\ell}=FP_{i}^{\ell}F^T+Q$. Besides, the conditional birth density $p^\ell_B(\tilde{x}|d)$ is also in the form of GM given by $\sum_{i=1}^{J_B(\ell)} w_{B,i}^{\ell,\tilde{x}} \mathcal{N}(\tilde{x} ; m_{B,i}^{\ell}, P_{B,i}^{\ell})$, which can be obtained by prior information or measurement driven birth model \cite{BDBB2012}. \\
\indent The update step begins with the computation of the normalization term $\eta _{Z}^{\left( \ell \right)}\left( x;\theta \right)$, which reads
\begin{equation}
    \begin{aligned}\label{eqn_etaZ}
        \eta _{Z}^{\left( \ell \right)}\left( x;\theta \right)&=\iint p^\ell_+(\tilde{x}|d')p^\ell_+(d)\frac{P_{D}^{\kappa}\left(d, \tau\right)g(\tilde{z},a|\tilde{x},d)}{\varphi(\tilde{z})c^\kappa_\tau(a)}\mathrm{d}\tilde{x}\mathrm{d}d\\
        &= \frac{1}{\varphi(\tilde{z})c^\kappa_\tau(a)}\int P_{D}^{\kappa}\left(d, \tau\right)p^\ell_+(d)p^\kappa_{\tau}(a|d)\mathrm{d}d\\
        &\quad \times \int p^\ell_+(\tilde{x}|d')g_{\tilde{z}}(\tilde{z}|\tilde{x})\mathrm{d}\tilde{x}\\
        &= \frac{1}{\varphi(\tilde{z})c^\kappa_\tau(a)}\zeta^\ell_Z(d;\theta)\sum_{i=1}^{J(\ell)}w_{i,+}^{\ell,\tilde{x}} \xi_i^\ell(\tilde{z})
    \end{aligned}
\end{equation}
for $\theta(\ell)>0$, where $\zeta^\ell_Z(d;\theta)=\int P_{D}^{\kappa}\left(d, \tau\right)p^\ell_+(d)p^\kappa_{\tau}(a|d)\mathrm{d}d$ and $\xi_i^\ell(\tilde{z})=\mathcal{N}(\tilde{z};Hm_{+,i}^{\ell},S^\ell_{i})$ with $S^\ell_{i}=HP_{+,i}^{\ell}H^T+R$. Then, an approximation of $p^\ell(\tilde{x}|d)$ based on \eqref{eqn_state_posterior} is given by
\begin{equation}\label{eqn_x_update}
    \begin{aligned}
        &~p^{\left( \theta \right)}\left( \tilde{x},\ell|d \right)\\
        &\approx \frac{P_{D}^{\kappa}\left(d, \tau\right)p^\ell_+(d)p^\kappa_{\tau}(a|d)}{\eta _{Z}^{\left( \ell \right)}\left( x;\theta\right)\varphi(\tilde{z})c^\kappa_\tau(a)}\sum_{i=1}^{J(\ell)}w_{i,+}^{\ell,\tilde{x}}\xi_i^\ell(\tilde{z})\mathcal{N}(\tilde{x};m_{i}^{\ell},P^\ell_{i})\\
        &=\sum_{i=1}^{J(\ell)}w_{i}^{\ell,\tilde{x}}\mathcal{N}(\tilde{x};m_{i}^{\ell},P^\ell_{i})
    \end{aligned}
\end{equation}
where $w_i^{\ell,\tilde{x}}=w_{i,+}^{\ell,\tilde{x}}\xi_i^\ell(\tilde{z})/\sum_{i=1}^{N(\ell)}w_{i,+}^{\ell,\tilde{x}}\xi_i^\ell(\tilde{z})$, $m_{i}^{\ell}=m_{+,i}^{\ell}+K^\ell_i\left(\tilde{z}-Hm_{+,i}^\ell\right)$, $K^\ell_i=P^\ell_{+,i}H^T\left(S^\ell_{i}\right)^{-1}$ and $P_i^\ell=P^\ell_{+,i}-P^\ell_{+,i}H^T\left(S^\ell_{i}\right)^{-1}HP^\ell_{+,i}$. \\
\indent For $\theta(\ell)=0$, i.e., target $\ell$ is miss detected, we have $\eta _{Z}^{\left( \theta \right)}\left( \ell \right)\approx 1-P_{D}^{\kappa}\left(d, \tau\right)$ and the updated Gaussian components are just the predicted ones. Given the association hypothesis set $\{(I_+,\theta)\}$, the posterior spatial density $p^\ell(\tilde{x})$ can be calculated by substituting $x$ with $\tilde{x}$ in \eqref{eqn_p_post}. For mildly nonlinear Gaussian dynamical and observation models, the GM realization of $p^\ell(\tilde{x}|d)$ can be obtained with the extended and unscented Kalman filters. In addition, the truncation procedure in \cite{VM2006} can be employed to relieve the problem that the number of Gaussian components increases greatly as time progresses. 
\section{SNR Estimation} \label{sec_four}
\indent Section \ref{sec_gm} presents the GM implementation of the spatial density conditioned on the target SNR. This section describes two methods for estimating the target SNR. One is the conventional SMC method for nonlinear and non-Gaussian filtering and the other is a Gamma estimator which approximates the density of the target SNR by Gamma distribution. We will put emphasis on the Gamma realization since it has the potential for avoiding the degeneracy problem with SMC method and reducing the computational load.
\subsection{SMC Method} \label{sec_d_smc}
\indent The SMC method is a well-known technique for Bayesian estimations that involve elements of nonlinearity and non-Gaussianity \cite{MSNT2002}. Thus, we first make use of the SMC method to estimate the target SNR, of which the process model \eqref{eqn_ncgamma} and the measurement likelihood \eqref{eqn_amplitude_lh} are both strong nonlinear functions. Using the SMC representation, the posterior density $p^\ell(d)$ can be denoted by a set of weighted particles
\begin{equation}\nonumber
    p^\ell(d)=\sum_{i=1}^{N(\ell)}w^{\ell,d}_{i}\delta_{d^\ell_i}(d).
\end{equation}
Under GM implementation of the conditional density $p^\ell(\tilde{x}|d)$ and SMC realization of the marginal density $p^\ell(d)$, the posterior $p^\ell(\tilde{x},d)=\sum_{i=1}^{N(\ell)}w^{\ell,d}_i p^\ell(\tilde{x}|d^\ell_i)\delta_{d^\ell_i}(d)$ can be denoted by $\{w^{\ell,d}_i,d^\ell_i,\mathcal{G}^\ell_i \}_{i=1}^{N(\ell)}$, where $\mathcal{G}^\ell_i=\{w^{\ell,\tilde{x}}_{i,j},m^\ell_{i,j},P^\ell_{i,j} \}_{j=1}^{N_j}$ is the parameter set of the GM representation of $p^\ell(\tilde{x}|d^\ell_i)$. Given a particle $d^\ell_{i}$, $i=1,...,N(\ell)$, the Gaussian components $\mathcal{G}^\ell_i$ attached to it can be computed according to Section \ref{sec_gm}.\\
\indent In the prediction step of target SNR, the NCG distribution \eqref{eqn_ncgamma} is chosen as the proposal, such that the predicted particles $d^\ell_{+,i}$ can be drawn directly from $f_d(d | d^\ell_{i};\delta,c,\rho)$ with weights $w^{\ell,d}_{+,i}=w^{\ell,d}_{i}$. The initial particles ia sampled uniformly from a possible SNR interval $[d_l,d_u]$ in dB domain, where $d_l$ and $d_u$ are the lower and upper bounds of the region. Then the predicted SNR density is given by 
\begin{equation}\label{eqn_smc_snr_predict}
    p^\ell_+(d)=\sum_{i=1}^{N(\ell)}w^{\ell,d}_{+,i}\delta_{d^\ell_{+,i}}(d).
\end{equation}
Substituting \eqref{eqn_smc_snr_predict} into \eqref{eqn_etaZ}, we have
\begin{equation}\label{eqn_smc_etaZ}
    \begin{aligned}
        \eta _{Z}^{\left( \ell \right)}\left( x;\theta \right)=&\frac{1}{\varphi(\tilde{z})c^\kappa_\tau(a)}\sum_{i=1}^{N(\ell)}w^{\ell,d}_{+,i} P_{D}^{\kappa}(d^\ell_{+,i}, \tau)\\
        &\times p^\kappa_{\tau}(a|d^\ell_{+,i})\sum_{j=1}^{N^i_j}w_{i,j}^{\ell,\tilde{x}} \xi_{i,j}^\ell(\tilde{z}).\\
    \end{aligned}
\end{equation}
Then measurement to track association hypotheses can be ranked according to the cost matrix made up of $\eta _{Z}^{\left( \ell \right)}\left( x;\theta \right)$. Given an association hypothesis $\theta$, the posterior SNR density of track $\ell$ is represented by
\begin{equation}\label{eqn_smc_snr_update}
    p^{(\theta)}(d,\ell)=\sum_{i=1}^{N(\ell)}w^{\ell,d}_{i}\delta_{d^\ell_{+,i}}(d),
\end{equation}
with weight $w^{\ell,d}_{i}=w^{\ell,d}_{+,i}p^\kappa_{\tau}(a|d^\ell_{+,i})/\sum_{i=1}^{N(\ell)}w^{\ell,d}_{+,i}p^\kappa_{\tau}(a|d^\ell_{+,i})$. \\
\indent A typical problem with the SMC method is the particle degeneracy phenomenon \cite{MSNT2002}, where after a few iterations, all but a few particles will have negligible weight. In order to reduce the effect of degeneracy, one can implement the resampling procedure on $\{w^{\ell,d}_i,d^\ell_i,\mathcal{G}^\ell_i \}_{i=1}^{N(\ell)}$. Besides, since the conditional kinematic state has to be calculated for every particle, the SMC method will result in great computational cost even for a few particles. Next, we propose a Gamma recursion of the target SNR to avoid or mitigate these problems. 
\subsection{Gamma Approximation} \label{sec_d_gamma}
\indent In \cite{M2019}, a Gamma estimator is developed for tracking local average RCS under the assumptions that the evolution of hidden state follows the NCG distribution, and the prior density of state and the measurement model follow a Gamma distribution. Inspired by this work, we attempt to derive a Gamma recursion of the target SNR density using amplitude measurement. Since the target SNR is strictly positive, suppose that the posterior density $p^\ell(d)$ corresponds to the Gamma distribution
\begin{equation}\label{eqn_snr_gamma}
    p^\ell(d) =\gamma \left( d;\alpha,\beta \right) =
    \begin{cases}
        0,&d\leq 0\\
        \frac{\beta^\alpha}{\Gamma (\alpha)}d^{\alpha-1}e^{-\beta d},& d>0
    \end{cases}
\end{equation}
characterized by shape $\alpha$ and rate $\beta$. Given the GM form of the spatial density $p^\ell(\tilde{x}|d)=\sum_{i=1}^{J(\ell)} w_{i}^{\ell,\tilde{x}} \mathcal{N}\left(\tilde{x} ; m_{i}^{\ell}, P_{i}^{\ell}\right)$, the joint density is $p^\ell(\tilde{x},d)=\gamma \left( d;\alpha,\beta \right)\sum_{i=1}^{J(\ell)} w_{i}^{\ell,\tilde{x}} \mathcal{N}\left(\tilde{x} ; m_{i}^{\ell}, P_{i}^{\ell}\right)$, which can be denoted by a triplet $(\alpha,\beta,\mathcal{G}^\ell)$ with Gaussian components $\mathcal{G}^\ell=\{w^{\ell,\tilde{x}}_{i},m^\ell_{i},P^\ell_{i} \}_{i=1}^{J(\ell)}$. The recursion of $\mathcal{G}^\ell$ has been described in Section \ref{sec_gm}. The procedure of estimating parameters $\alpha$ and $\beta$ is presented as follows.
\subsubsection{Time Update} 
\begin{figure*}[t]
    \centering
    \subfloat[$\beta=0.5$]{
        \includegraphics[width=2.2in]{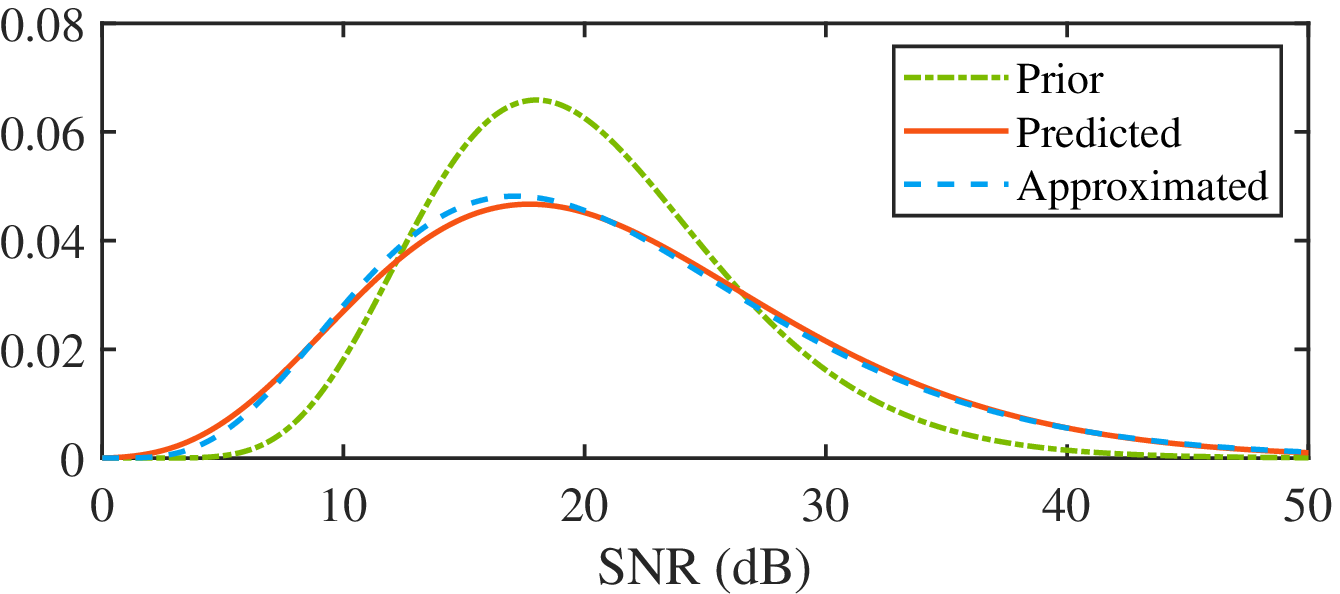}
    }
    \subfloat[$\beta=1$]{
        \includegraphics[width=2.2in]{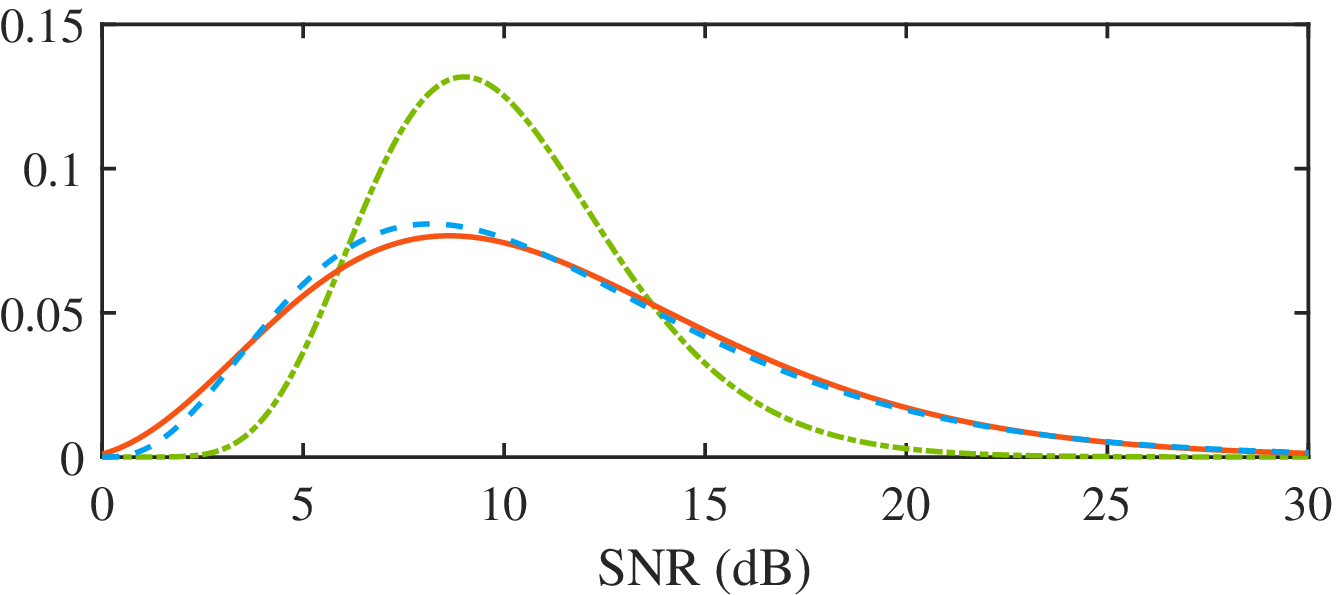}
    }
    \subfloat[$\beta=3$]{
        \includegraphics[width=2.2in]{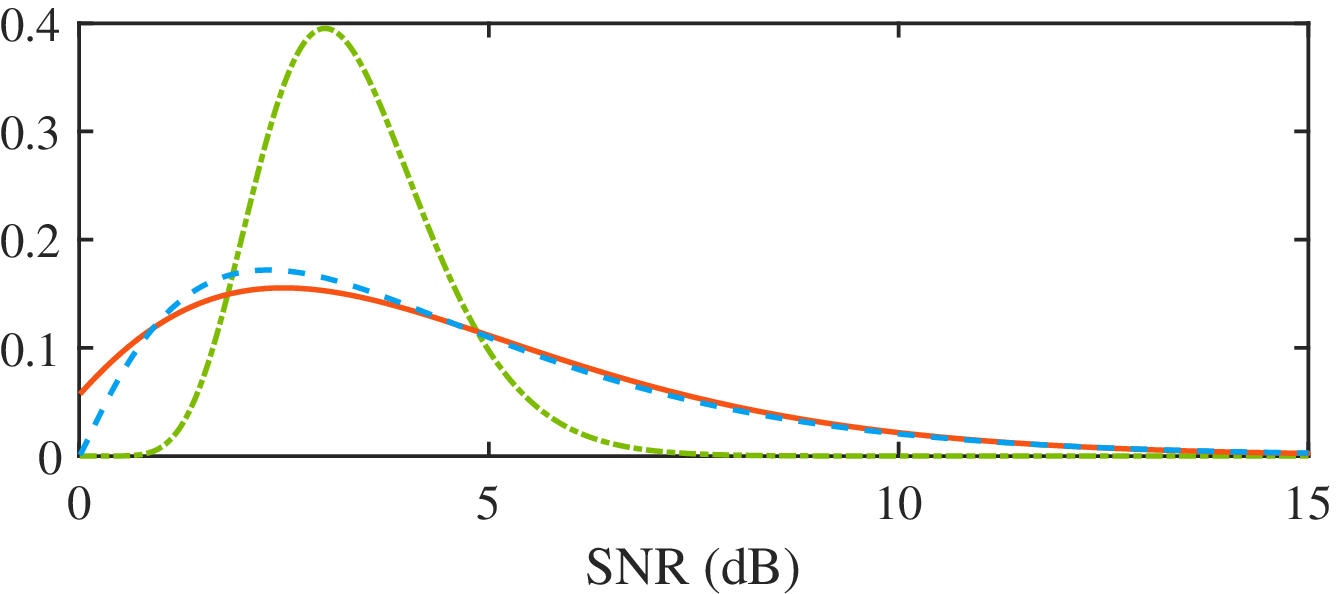}
    }
    \caption{Comparisons of prior, predicted and approximate SNR densities for $c=1,\delta=1~ (\text{Swerling 1 target}),\rho=0.999, \alpha=10$ and three choices of $\beta$.}
    \label{fig_snr_predict_d1}
    \vspace{-4mm}
\end{figure*}

\begin{figure*}[t]
    \centering
    \subfloat[$\beta=0.5$]{
        \includegraphics[width=2.2in]{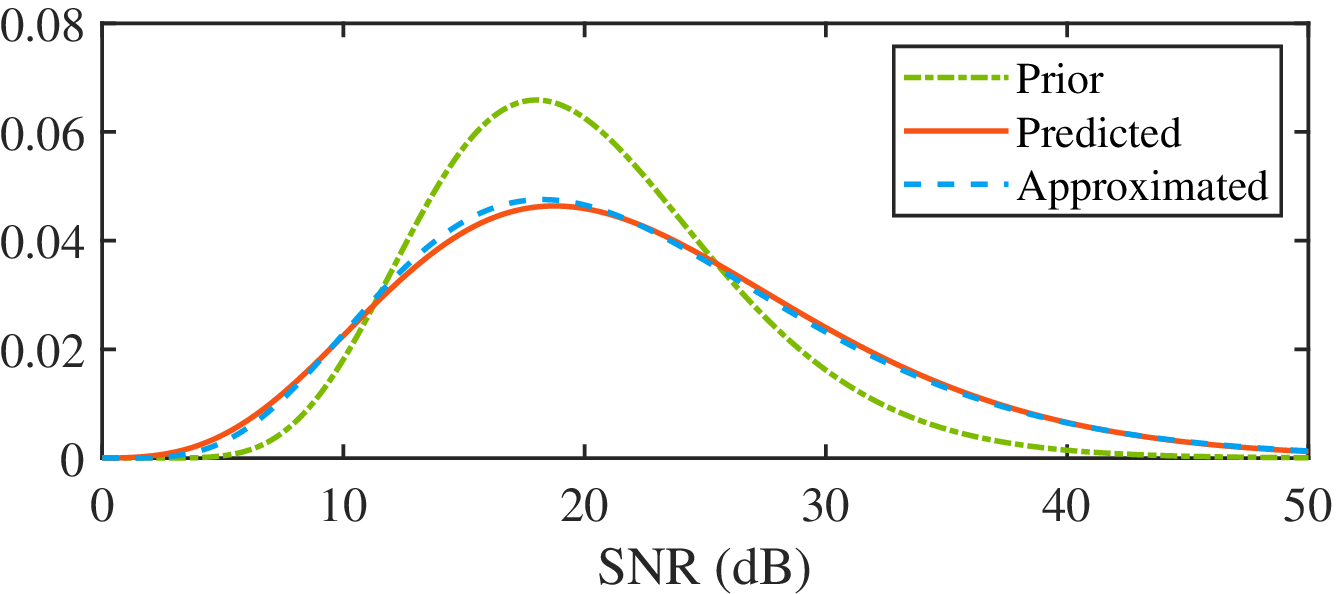}
    }
    \subfloat[$\beta=1$]{
        \includegraphics[width=2.2in]{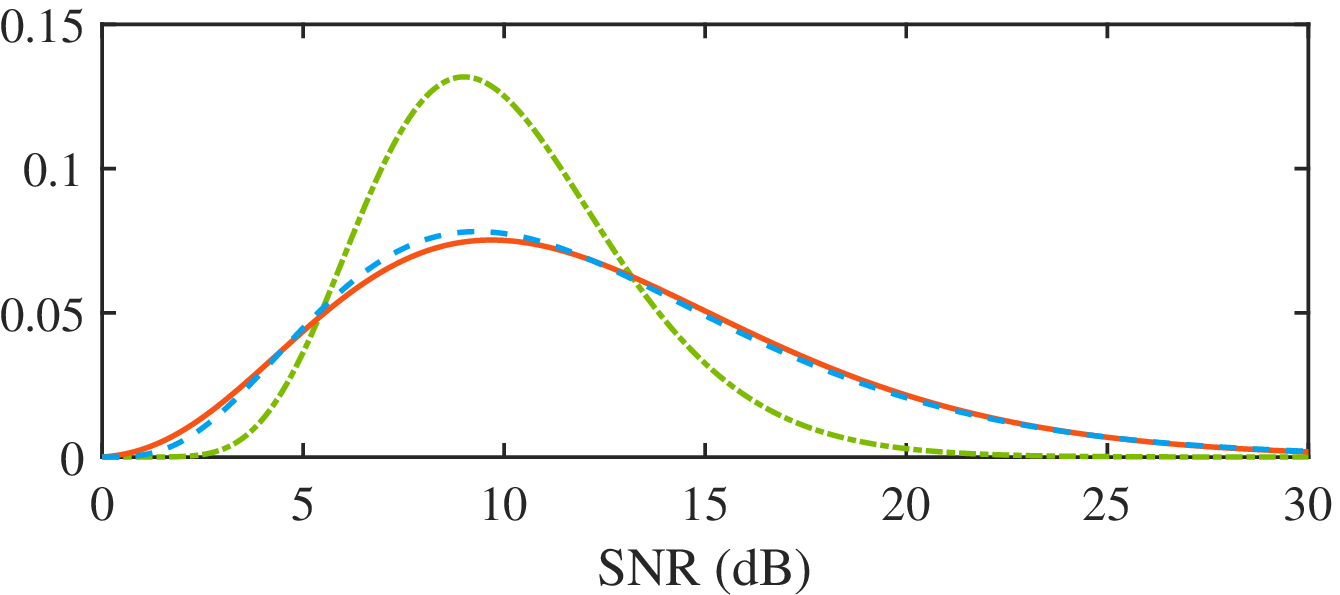}
    }
    \subfloat[$\beta=3$]{
        \includegraphics[width=2.2in]{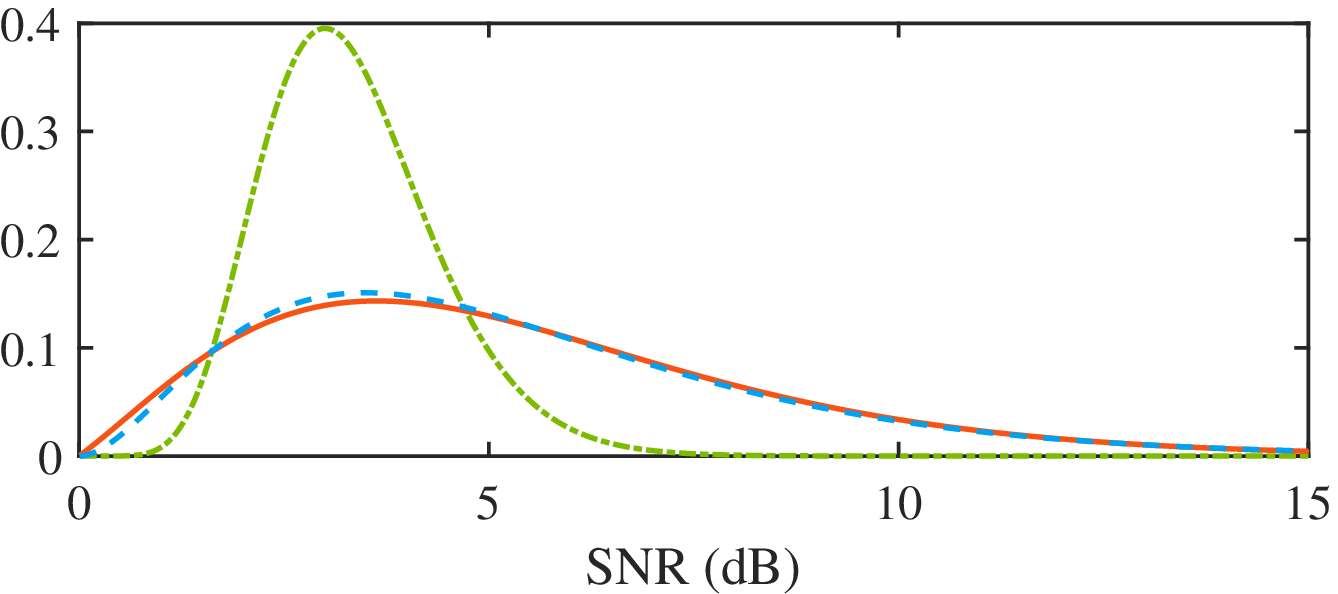}
    }
    \caption{Comparisons of prior, predicted and approximate SNR densities for $c=1,\delta=2~ (\text{Swerling 3 target}),\rho=0.999, \alpha=10$ and three choices of $\beta$.}
    \label{fig_snr_predict_d2}
\end{figure*}

The predictive SNR density given by \eqref{eqn_snr_predict} is intractable since the SNR transition density $f_d(d | d')$, i.e., the NCG distribution \eqref{eqn_ncgamma}, involves a summation over infinite elements. Nevertheless, the Laplace transform of $f_d(d | d')$ is remarkably simple \cite{M2019}
\begin{equation}\label{eqn_laplace_snr_trans}
    \mathcal{L}(f_d(d | d';\delta,\rho,c))=\frac{1}{(sc+1)^{\delta}}\exp \left( -\frac{s\rho d'}{sc+1} \right) .
\end{equation}
Substitution of \eqref{eqn_laplace_snr_trans} into the SNR prediction equation \eqref{eqn_snr_predict}, based on inverse Laplace transform, leads to
\begin{equation} \label{eqn_d_invLaplace}
    \begin{aligned}
        &~p^\ell_+(d)\\
        &=\int_0^\infty f^\ell_d(d|d')p^\ell(d')\mathrm{d}d'\\
        &=\int_0^\infty p^\ell(d')\mathrm{d}d'\int^{\sigma+j\infty}_{\sigma-j\infty}\frac{e^{sd}}{(sc+1)^{\delta}}\exp \left( -\frac{s\rho d'}{sc+1} \right) \mathrm{d}s\\
        &=\int^{\sigma+j\infty}_{\sigma-j\infty}K(s)e^{sd}\mathrm{d}s\\
        &=\mathcal{L}^{-1}\left(K(s)\right),
    \end{aligned}
\end{equation}
where 
\begin{equation}\nonumber
K(s)=\int_0^\infty p^\ell(d')\frac{1}{(sc+1)^{\delta}}\exp \left( -\frac{s\rho d'}{sc+1} \right)\mathrm{d}d'.
\end{equation}
Inserting \eqref{eqn_snr_gamma} into $K(s)$, after a few steps, we have
\begin{equation}
K(s)=\frac{\left( sc+1 \right) ^{\alpha-\delta}}{\left( s(\rho\beta^{-1}+c) +1  \right) ^{\alpha }}.
\end{equation}
\indent Figs. \ref{fig_snr_predict_d1} and \ref{fig_snr_predict_d2} show the comparisons of the prior density $p^\ell(d)=\gamma(d;\alpha,\beta)$ with the corresponding predicted density $p^\ell_+(d)$ for Swerling 1 and 3 targets, given $c=1$, $\alpha=10$, $\rho=0.999$ and three choices of $\beta$. The predictive density $p^\ell_+(d)$ is obtained by computing the inverse Laplace transform of $K(s)$ numerically \cite{GU1983}. The shape of $p^\ell_+(d)$ is similar to that of a Gamma distribution, which indicates the possibility of approximating $p^\ell_+(d)$ by Gamma PDF.\\
\indent In fact, the Laplace transform of a Gamma distribution $\gamma ( d;\tilde{\alpha},\tilde{\beta} )$ reads 
\begin{equation}\nonumber
    K_{\gamma}(s)=\frac{1}{\left( s\tilde{\beta}^{-1}+1 \right) ^{\tilde{\alpha}}},
\end{equation}
which mainly differs from $K(s)$ by the absence of zeros or poles induced by $c$. However, since $c$ is a small constant, these zeros or poles of $K(s)$ are located in the high-frequency range and far away from the imaginary axis, such that their influence is negligible. Besides, the dominant poles of $K(s)$ approach to the poles of $K_{\gamma}(s)$ as $\rho\rightarrow 1$ and $c\rightarrow 0$. \\
\indent A simple Gamma approximation of $p^\ell_+(d)$ can be obtained using the moment matching technique. Let the first two moments of $p^\ell_+(d)$ be equal to the moments of its approximation
\begin{equation}\label{eqn_moment_matching}
    \begin{aligned}
        \left.\frac{\mathrm{d}}{\mathrm{d}s}K_\gamma(s)\right|_{s\rightarrow0}&=\left.\frac{\mathrm{d}}{\mathrm{d}s}K(s)\right|_{s\rightarrow0},\\
        \left.\frac{\mathrm{d}^2}{\mathrm{d}s^2}K_\gamma(s)\right|_{s\rightarrow0}&=\left.\frac{\mathrm{d}^2}{\mathrm{d}s^2}K(s)\right|_{s\rightarrow0},
    \end{aligned}
\end{equation}
where the mean and variance of $K_{\gamma}(s)$ are $-\tilde{\alpha}/\tilde{\beta}$ and $\tilde{\alpha}(\tilde{\alpha}+1)/\tilde{\beta}^2$, respectively \cite{M2019}. The terms on the right-hand side of \eqref{eqn_moment_matching} can be obtained after a few straightforward derivations
\begin{equation}\label{ewn_M1M2}
    \begin{aligned}
        M^1_K\triangleq&\left. \frac{\mathrm{d}}{\mathrm{d}s}K(s) \right|_{s\rightarrow 0}=-\delta c-\frac{\alpha}{\beta},\\
        M^2_K\triangleq&\left. \frac{\mathrm{d}^2}{\mathrm{d}s^2}K(s) \right|_{s\rightarrow 0}=\delta c^2\left( \delta +1 \right) +\frac{\rho^2\alpha\left( \alpha +1 \right)}{\beta ^{2}}\\
        &\qquad\qquad\qquad\quad+\frac{2\rho c\alpha \left( \delta +1 \right)}{\beta}.\\
    \end{aligned}
\end{equation}
Substituting \eqref{ewn_M1M2} into \eqref{eqn_moment_matching}, the parameters of the approximate density $p^\ell_+(d)\approx\gamma ( d;\tilde{\alpha},\tilde{\beta} )$ are given by
\begin{equation} \label{eqn_ab_predict}
    \tilde{\alpha}=\frac{(M^1_K)^{2}}{M^2_K-(M^1_K)^{2}},\quad \tilde{\beta}=\frac{M^1_K}{M^2_K-(M^1_K)^{2}}.
\end{equation}
\indent It can be seen from Figs. \ref{fig_snr_predict_d1} and \ref{fig_snr_predict_d2} that the approximate Gamma distributions well match the predicted SNR densities. Formally, the accuracy of the Gamma approximation can be measured with the Kullback–Leibler divergence (KLD) \cite{AMGC2005} between the true density \eqref{eqn_d_invLaplace} and the approximate density $\gamma ( d;\tilde{\alpha},\tilde{\beta} )$. The KLD is zero if and only if the two densities are identical, and a small value indicates that there is merely a slight difference between the two distributions and hence a reasonable matching. All the KLD values of the testing conditions in Figs. \ref{fig_snr_predict_d1} and \ref{fig_snr_predict_d2} are less than 0.02, suggesting that the Gamma distribution provides a good approximation of $p^\ell_+(d)$ for a fairly wide range of model parameters. 
\subsubsection{Measurement Update} 
\begin{figure} [t]
    \centering
    \subfloat[]{
        \includegraphics[width=1.65in]{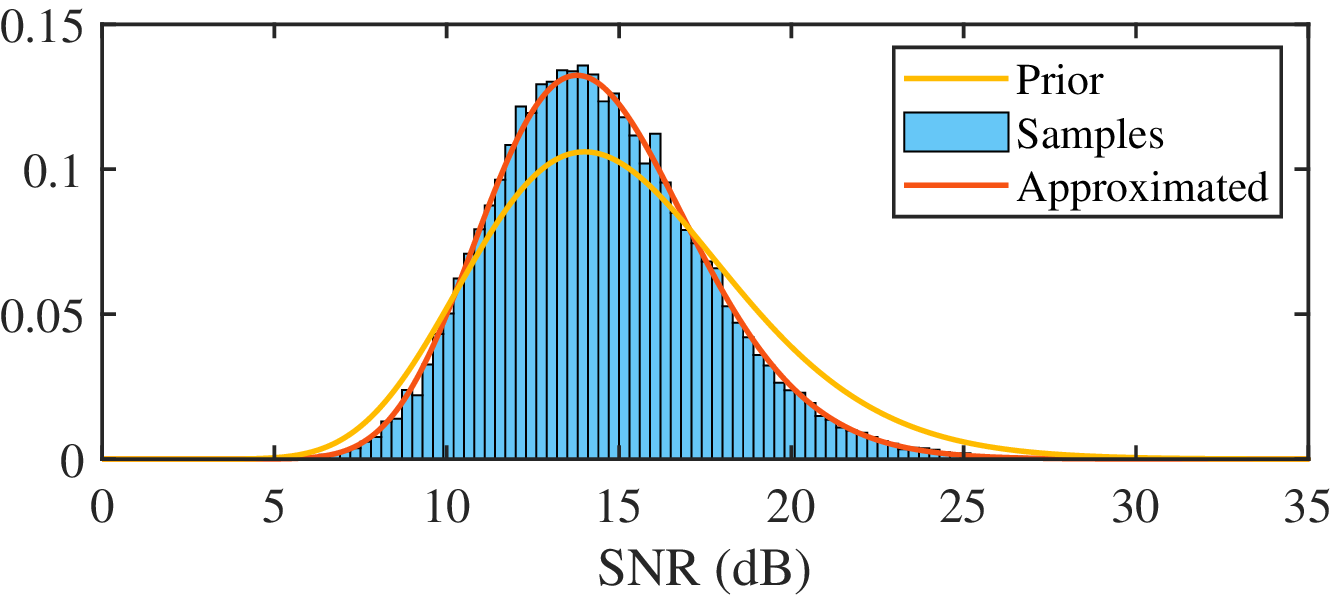}
    }\hspace{-3mm}
    \subfloat[]{
        \includegraphics[width=1.65in]{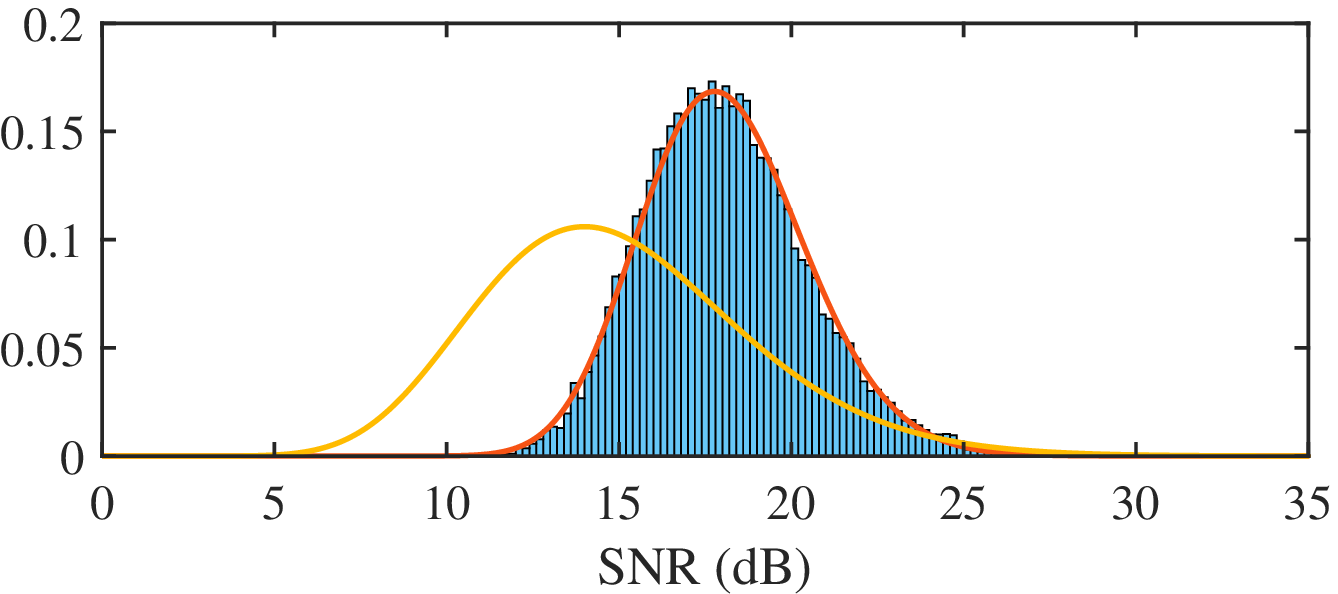}
    }
    \caption{Approximate posterior densities of the target SNR for $\tilde{\alpha}=10,\tilde{\beta}=1,\tau=2,N_m=50000$. (a) Swerling 1 amplitude likelihood given $a=6.9$. (b) Swerling 3 amplitude likelihood given $a=10.6$.}
    \label{fig_snr_update_mhs1}
\end{figure}

Inserting the predictive density $p^\ell_+(d)=\gamma(d;\tilde{\alpha},\tilde{\beta})$ and the amplitude likelihood $p^\kappa_{\tau}(a|d)$ into the SNR update formula \eqref{eqn_snr_update}, we have
\begin{equation} \label{eqn_snr_posterior1}
    p^{(\theta)}(d,\ell)=\frac{1}{\psi_1}\frac{a}{1+d}\exp\left(\frac{\tau^2-a^2}{2\left(1+d\right)}\right)\frac{\tilde{\beta}^{\tilde{\alpha}}d^{\tilde{\alpha}-1}}{\Gamma(\tilde{\alpha})}e^{-\tilde{\beta}d},
\end{equation}
\begin{equation} \label{eqn_snr_posterior3}
    \begin{aligned}
    p^{(\theta)}(d,\ell)=&\frac{1}{\psi_3}\frac{9a^3}{3\tau^2(1+d)+2(1+d)^2}\\
    &\times\exp\left(\frac{3(\tau^2-a^2)}{2\left(1+d\right)} \right)\frac{\tilde{\beta}^{\tilde{\alpha}}d^{\tilde{\alpha}-1}}{\Gamma(\tilde{\alpha})}e^{-\tilde{\beta}d},\\
    \end{aligned}
\end{equation}
for Swerling 1 and 3 amplitude likelihoods, where $\psi_1$ and $\psi_3$ are the normalizing constants cannot be computed analytically. It can be seen that either \eqref{eqn_snr_posterior1} or \eqref{eqn_snr_posterior3} is too complicated for the next iteration, thus a simple approximation is needed.\\
\indent The existing solutions to the approximation of complex posterior can be roughly divided into deterministic approaches, such as Laplace approximation and variational Bayes, and stochastic approaches, such as importance sampling, accept-reject, MCMC \cite{GCSDVR2013,AA1990}. Deterministic methods usually involve derivations that may be intractable for \eqref{eqn_snr_posterior1} and \eqref{eqn_snr_posterior3} or require the knowledge of the exact form of the approximated distribution, hence we consider the stochastic methods. In particular, we employ the MCMC method to reveal the shape of the posterior density by the histogram of samples. An MCMC algorithm generates an ergodic Markov chain $\mathcal{M}$ with the density of interest being its stationary distribution, such that the chain converges to the target density asymptotically. Since our goal is to sample from a univariate distribution, we employ the classical Metropolis-Hastings (MH) \cite{W1970} algorithm to construct the Markov chain $\mathcal{M}$. For high-dimensional problems, Gibbs \cite{GG1984} or Hamiltonian Monte Carlo \cite{GC2011} samplers are recommended. \\
\indent To generate samples from the target density $p^{(\theta)}(d,\ell)$ by the MH algorithm, suppose that the chain $\mathcal{M}$ is currently at state $d_i$. Then the chain jumps to a new state $d_{i+1}$ according to the proposal $q(d_{i+1}|d_i)$ with probability 
\begin{equation}\label{eqn_hastings_ratio}
    A(d_i,d_{i+1})=\min\left(1,\frac{p^{(\theta)}(d_{i+1},\ell)q(d_i|d_{i+1})}{p^{(\theta)}(d_{i},\ell)q(d_{i+1}|d_{i})}\right),
\end{equation}
otherwise the chain stays at $d_i$. The sampler always accepts $d_{i+1}$ such that the ratio $p^{(\theta)}(d_{i+1},\ell)/q(d_{i+1}|d_{i})$ increases compared with the previous value $p^{(\theta)}(d_{i},\ell)/q(d_{i}|d_{i+1})$, but it may accept values $d_{i+1}$ such that the ratio decreases. This behavior ensures the ergodic of the Markov chain. It can be observed from \eqref{eqn_hastings_ratio} that the normalizing constant is offset due to the ratio $p^{(\theta)}(d_{i+1},\ell)/p^{(\theta)}(d_{i},\ell)$, which facilitates us to draw samples from the density known only up to a constant. The sampling procedure is summarized in Algorithm \ref{alg_mh}, where $N_m$ is the number of samples. \\
\begin{algorithm}[t]  
    \caption{The sampling procedure based on MH}  \label{alg_mh}
    \begin{algorithmic}[1]
        \small
        \Require $\tilde{\alpha},\tilde{\beta},a,N_m$ 
        \Ensure $\{d_i\}_{i=1}^{N_m}$
        \State Draw the initial sample $d_1\sim p^\ell_+(d)=\gamma(d;\tilde{\alpha},\tilde{\beta})$ 
        \For {$i = 1,...,N_m-1$}
        \State Draw a sample $d'\sim q(d|d_{i})$ 
        \State Accept the new state, $d_{i+1}=d'$, with probability $A(d_i,d')$ 
        \Statex \quad ~ given by \eqref{eqn_hastings_ratio}. Otherwise, $d_{i+1}=d_{i}$
        \EndFor
    \end{algorithmic}  
\end{algorithm} 
\indent Two examples are given here to illustrate the effectiveness of the MH algorithm for revealing the shape of the density $p^{(\theta)}(d,\ell)$. Suppose that the predicted SNR density is approximated by $p^\ell_+(d)=\gamma(d;10,1)$. Given the detection threshold $\tau=2$, the amplitude measurements for Swerling 1 and 3 targets are generated by the inverse method and acceptance-rejection method \cite{D1986}, respectively. A Gaussian distribution $p(d_{i+1}|d_i)=\mathcal{N}(d;d_i;4^2)$ is chosen as the proposal to generate $50000$ samples. Fig. \ref{fig_snr_update_mhs1}(a) and Fig. \ref{fig_snr_update_mhs1}(b) show the normalized histograms of the simulated samples for Swerling 1 and 3 targets. It can be seen that the distributions of the samples are close to the Gamma distributions (red curves), which presents an opportunity to approximate $p^{(\theta)}(d,\ell)$ using Gamma PDF.\\
\indent Let the posterior SNR density be approximated by a Gamma distribution, i.e., $p^{(\theta)}(d,\ell)\approx\gamma(d;\alpha,\beta)$, where $\alpha$ and $\beta$ can be obtained by equalizing the first two moments of $\gamma(d;\alpha,\beta)$ and the corresponding moments of the generated samples. The mean and variance of $\gamma(d;\alpha,\beta)$ are equal to $\alpha/\beta$ and $\alpha/\beta^2$, while the mean and variance of the samples are given by
\begin{equation}\label{eqn_sample_mean_var}
    \bar{d}=\frac{1}{N_m}\sum_{i=1}^{N_m}d_i,\quad S^2=\frac{1}{N_m-1}\sum_{i=1}^{N_m}\left(d_i-\bar{d}\right)^2.
\end{equation}
Then, based on moment matching, we have
\begin{equation}\label{eqn_ab_update}
    \alpha=\frac{\bar{d}^2}{S^2},\quad\beta=\frac{\bar{d}}{S^2}.
\end{equation}
\indent The red curves in Fig. \ref{fig_snr_update_mhs1} represent the approximated Gamma posteriors based on the samples. In theory, the distribution of the samples exactly represents the target density as the sample size $N_m$ tends to infinity. However, one has to choose an appropriate number of samples due to insufficient computational power or real-time constraint in practice. The sample size can be determined by the Gelman-Rubin (GR) diagnostic \cite{GR1992, GCSDVR2013}, which monitors the convergence of a MCMC sampler. \\
\indent As we approximate the predicted and updated SNR densities by Gamma distributions, the MMSE estimates of the corresponding SNRs read 
\begin{equation}
    \hat{d}_+=\int dp^\ell_+(d)\mathrm{d}d=\frac{\tilde{\alpha}}{\tilde{\beta}},\quad \hat{d}=\int dp^{(\theta)}(d,\ell)\mathrm{d}d=\frac{\alpha}{\beta},
\end{equation}
where $\tilde{\alpha}$ and $\tilde{\beta}$, $\alpha$ and $\beta$ are given by \eqref{eqn_ab_predict} and \eqref{eqn_ab_update}. Moreover, the likelihood $\eta _{Z}^{\left( \ell \right)}\left( x;\theta \right)$ in \eqref{eqn_etaZ} can be approximated with
\begin{equation}\label{eqn_gamma_etaZ}
    \begin{aligned}
        \eta _{Z}^{\left( \ell \right)}\left( x;\theta \right)&= \frac{\sum_{i=1}^{J(\ell)}w_i^{\ell,\tilde{x}} \xi_i^\ell(\tilde{z})}{\varphi(\tilde{z})c^\kappa_\tau(a)}\int P_{D}^{\kappa}(d, \tau)p^\ell_+(d)p^\kappa_{\tau}(a|d)\mathrm{d}d\\
        &\approx \frac{\sum_{i=1}^{J(\ell)}w_i^{\ell,\tilde{x}} \xi_i^\ell(\tilde{z})}{\varphi(\tilde{z})c^\kappa_\tau(a)}P_{D}^{\kappa}(\hat{d}_+, \tau)p^\kappa_{\tau}(a|\hat{d}_+),
    \end{aligned}
\end{equation}
then the measurement to track association hypotheses can be extracted based on the cost matrix. Since the update step \eqref{eqn_p_post} contains multiple association hypothesis, the posterior SNR density $p^\ell(d)$ is actually a mixture of Gamma distributions, which is intractable for further recursion and truncation. For simplicity, we use the density $p^{(\theta)}(d,\ell)$ with maximum weight $w^{(I_+,\theta)}(Z)$ to approximate $p^\ell(d)$, i.e., 
\begin{equation}
    p^\ell(d)\approx p^{(\theta_*)}(d,\ell),~ \theta_*=\arg\max_{\theta}w^{(I_+,\theta)}(Z).
\end{equation}
With Gamma approximation of $p^\ell(d)$, the kinematic state is simply conditioned on the estimated SNR instead of each particle as in the SMC realization, which provides an opportunity to reduce computational load. 
\subsection{A summary of the algorithm} \label{sec_summary}
\indent A complete implementation of the proposed HLMB filter can be obtained by combining the GM representation of the kinematic state $p^\ell(\tilde{x}|d)$ with SMC or Gamma approximation of the SNR density $p^\ell(d)$. Denote by GM/SMC-HLMB the hybrid filter with GM and SMC implementations, and by GM/G-HLMB with GM and Gamma. Since the density of the augmented state is decomposed as $p^{\ell}(\tilde{x},d)=p^\ell(\tilde{x}|d)p^\ell(d)$, then the LMB RFSs for the existing and the newborn tracks can be represented by the parameter sets $\left\{(r^{\ell},p^\ell(\tilde{x}|d)p^\ell(d))\right\}_{\ell\in I}$ and $ \left\{(r_B^\ell,p_B^\ell(\tilde{x}|d)p_B^\ell(d))\right\}_{\ell\in I_B}$, where $I$ and $I_B$ are the corresponding label sets. Algorithm \ref{alg_hlmb} summarizes the proposed HLMB filter. \\
\indent The GM/SMC-HLMB filter is a multi-target extension of the single-target Bernoulli filter with amplitude proposed in \cite{RRKG2021}, thus it can be regarded as a comparative method. In the next section, we will compare the proposed GM/G-HLMB filter with the GM/SMC-HLMB, the GM-LMB without amplitude, the GM-LMB with marginalized amplitude likelihood (GM-LMB-M), as well as the GM-LMB with amplitude and known target SNR (GM-LMB-K). The GM-LMB-M and GM-LMB-K filters avoid the estimation of the target SNR, and only utilize target amplitude to modify the measurement likelihood \eqref{eqn_meas_lh} with either marginalized amplitude likelihood \cite{CRVV2010} or amplitude likelihood with known SNR. \\
\begin{algorithm}[htbp]  
    \caption{A cycle of the HLMB filter}  \label{alg_hlmb}
    \begin{algorithmic}[1]
        \small
        \Require Existing track $\left\{(r^{\ell},p^\ell(\tilde{x}|d)p^\ell(d))\right\}_{\ell\in I}$ and newborn track $ \left\{(r_B^\ell,p_B^\ell(\tilde{x}|d)p_B^\ell(d))\right\}_{\ell\in I_B}$ at time $k-1$, measurement set $Z_k$ 
        \Ensure Existing track $\left\{(r^{\ell},p^\ell(\tilde{x}|d)p^\ell(d))\right\}_{\ell\in I}$ and newborn track $\left\{(r_B^\ell,p_B^\ell(\tilde{x}|d)p_B^\ell(d))\right\}_{\ell\in I_B}$ at time $k$
        \For {$i=1,...,|I|$}
        \State Compute $r^{\ell_i}_+$ and $p^{\ell_i}_+(\tilde{x}|d)$ according to \eqref{eqn_r_predict} and \eqref{eqn_x_predict}
        \State Compute  $p^{\ell_i}_+(d)$ according to \eqref{eqn_smc_snr_predict}/\eqref{eqn_ab_predict}
        \EndFor
        \State $\boldsymbol{\pi}_+=\left\{(r^\ell_+,p^\ell(\tilde{x}|d)p^\ell_+(d))\right\}_{\ell\in I}\cup\left\{(r_B^\ell,p_B^\ell(\tilde{x}|d)p_B^\ell(d))\right\}_{\ell\in I_B}$ with label set $I_+=I\cup I_B$
        \State Construct the cost matrix $C$ with $\eta _{Z}^{\left( \ell \right)}\left( x;\theta \right)$ given by \eqref{eqn_smc_etaZ}/\eqref{eqn_gamma_etaZ} 
        \State Extract association hypotheses $\{(I_+,\theta)\}_{\theta\in\Theta_{I_+}}$ from $C$ 
        \For {$i=1,...,|I_+|$}
            \For {$j=1,...,|\Theta_{I_+}|$}
            \If{$\theta_j(\ell_i)>0$}
            \State {Compute $p^{(\theta_j)}(\tilde{x},\ell_i|d)$ according to \eqref{eqn_x_update} 
            \State Compute $p^{(\theta_j)}(d,\ell_i)$ according to \eqref{eqn_smc_snr_update}}/\eqref{eqn_ab_update}
            \Else
            \State Let $p^{(\theta_j)}(\tilde{x},\ell_i|d)$ and $p^{(\theta_j)}(d,\ell_i)$ be the predicted 
            \Statex \qquad\qquad~densities
            \EndIf
            \EndFor
        \EndFor
        \For {$i=1,...,|I_+|$}
        \State Compute $r^{\ell_i}$ according to \eqref{eqn_r_post} 
        \State Compute $p^{\ell_i}(\tilde{x}|d)$ and $p^{\ell_i}(d)$ according to \eqref{eqn_p_post} 
        \EndFor 
        \State Let $I=I_+$, then we have $\left\{(r^{\ell},p^\ell(\tilde{x}|d)p^\ell(d))\right\}_{\ell\in I}$
        \State Obtain $\left\{(r_B^\ell,p_B^\ell(\tilde{x}|d)p_B^\ell(d))\right\}_{\ell\in I_B}$ from $Z_k$
    \end{algorithmic}  
\end{algorithm}   
\section{Simulation}\label{sec_five}
\subsection{Setup}
\begin{figure}[t]
    \centering
    \subfloat[Position]{
        \includegraphics[width=1.65in]{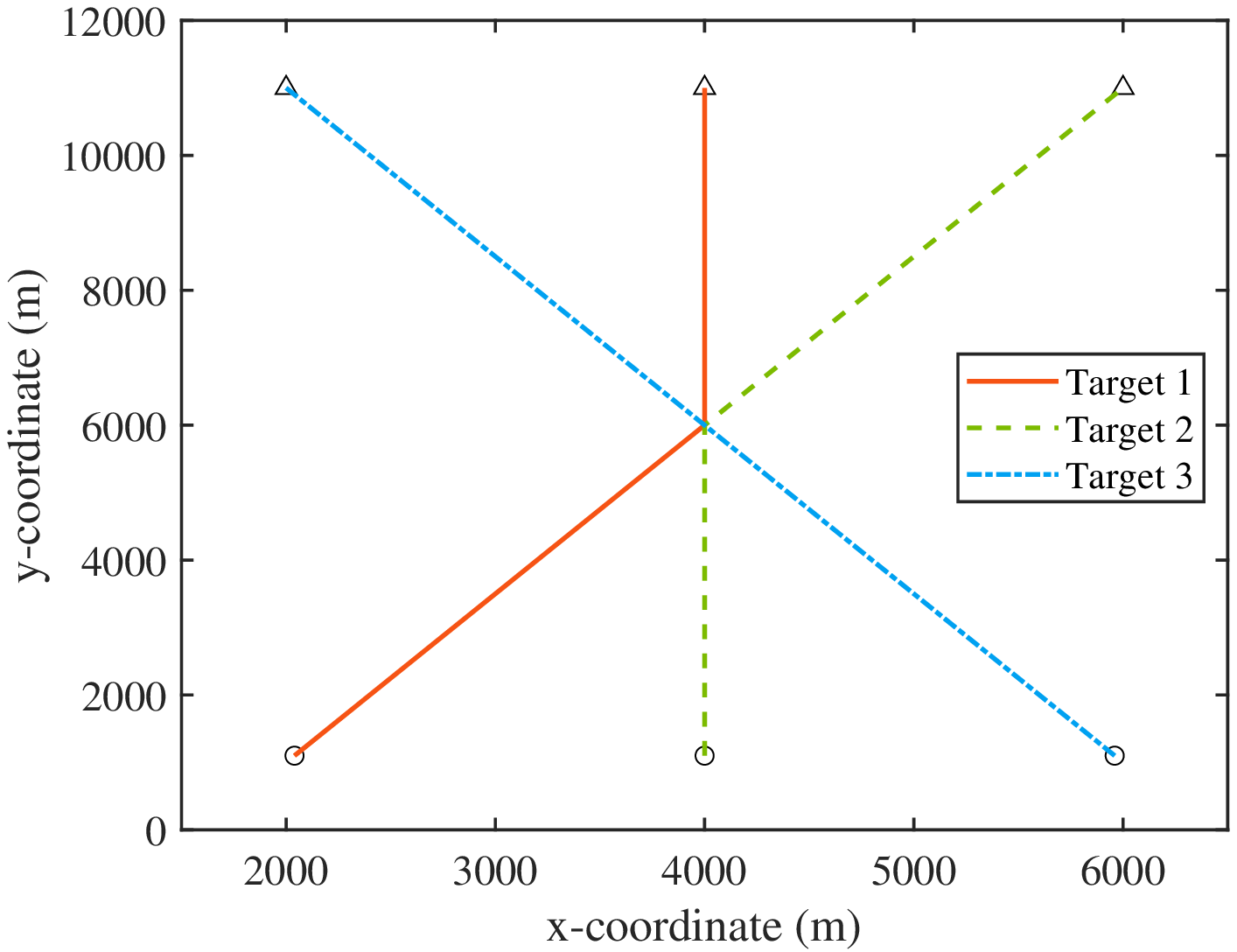}
        \label{fig_ground_truth_position}
    }\hspace{-3mm}
    \subfloat[SNR]{
        \includegraphics[width=1.65in]{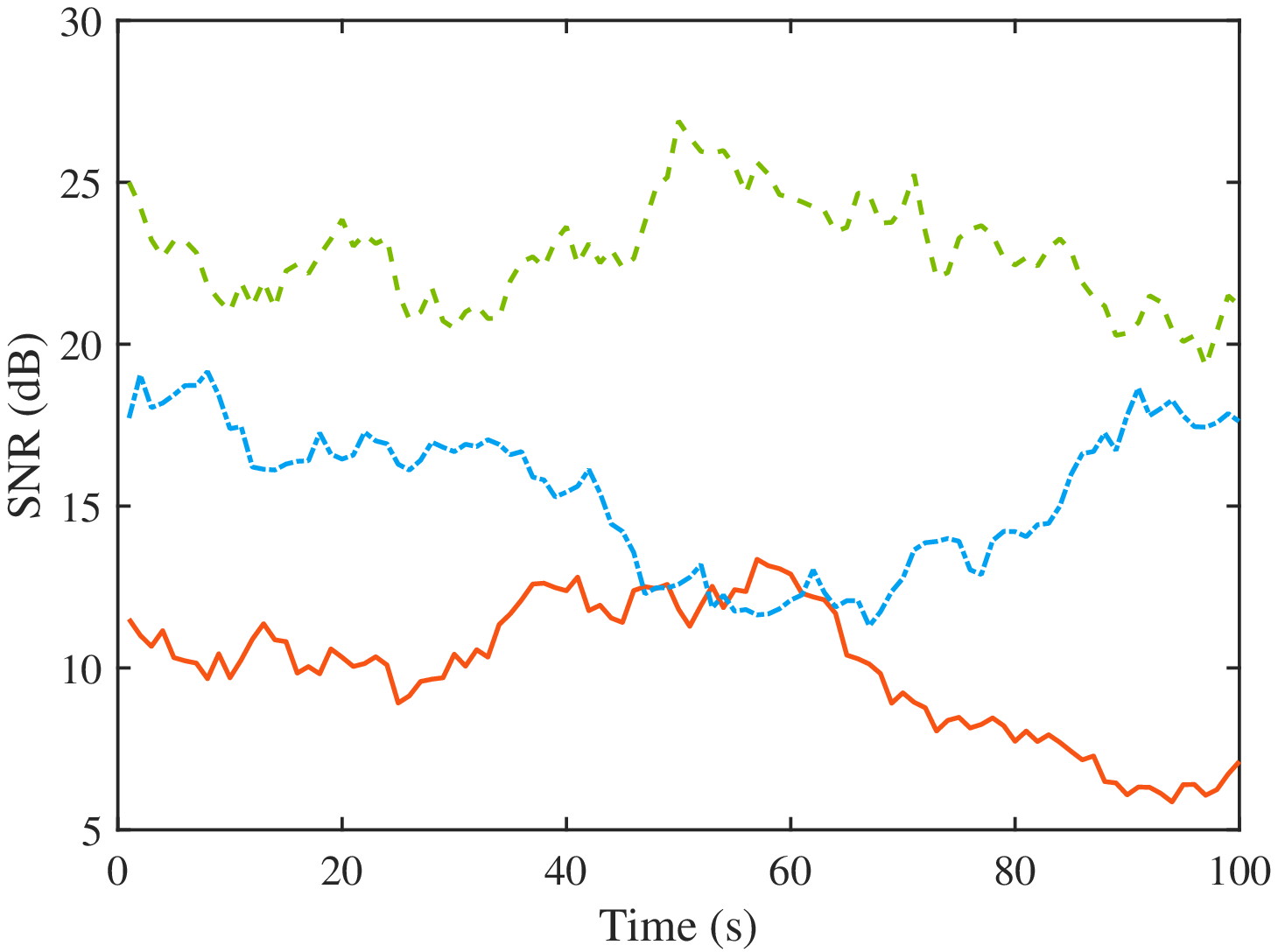}
    }
    \caption{Ground truth trajectories of the testing scenario. $\circ$ and $\triangle$ denote the start and stop positions.}
    \label{fig_ground_truth}
    \vspace{-3mm}
\end{figure}

\indent This section evaluates the performance of the proposed GM/G-HLMB filter compared with the methods mentioned in Section \ref{sec_summary}. The testing scenario consists of three crossing targets on a two dimensional region $[0,12000]\mathrm{m}$$\times[0,12000]\mathrm{m}$ as depicated in Fig. \ref{fig_ground_truth}(a), which is similar to the configuration in \cite{MUK2016}. The transition density for the kinematic state $\tilde{x}$ and the likelihood of the position measurement $\tilde{z}$ are both linear Gaussian functions given by $f_{\tilde{x}}(\tilde{x}|\tilde{x}')=\mathcal{N}(\tilde{x};F\tilde{x}',Q)$ and $g_{\tilde{z}}(\tilde{z}|\tilde{x})=\mathcal{N}(\tilde{z};H\tilde{x},R)$, respectively, with parameters
\begin{equation}\nonumber
    \begin{aligned}
        F&=I_2\otimes
        \begin{bmatrix}
            1 & \Delta \\
            0 & 1 \\
        \end{bmatrix},&\quad Q&=I_2\otimes
        \begin{bmatrix}
            \Delta^4/4 & \Delta^3/2 \\
            \Delta^3/2 & \Delta^2 \\
        \end{bmatrix}\sigma^2_v, \\
        H&=\begin{bmatrix}
            1 & 0 & 0 & 0 \\
            0 & 0 & 1 & 0 \\
        \end{bmatrix},&\quad R &= \sigma_\varepsilon^2I_2,
    \end{aligned}
\end{equation}
where $I_n$ is the $n\times n$ identity matrix, $\otimes$ denotes the Kronecker product, $\Delta=1$ s is the sampling period, $\sigma_v=10~\mathrm{m/s^2}$ and $\sigma_\varepsilon=20$ m are the standard derivations of process noise and observation noise. The duration of the simulation is $K=100$ s. Each target has the state independent survival probability $p_S= 0.98$.\\
\indent The three targets are born at $k = 1$ s and present throughout the simulation. The initial states of the three target are $x_0^1=[2000, 40, 1000, 100, 12]^T$, $x_0^2=[4000, 0, 1000, 100, 25]^T$ and $x_0^3=[6000, -40, 1000, 100, 17]^T$. The three targets travel along straight lines until they cross at $k=$50 s, then targets 1 and 2 change their velocities along the x-axis to 0 m/s and 40 m/s, hence changing their directions. This configuration will lead to incorrect measurement to track associations and track switchings at the intersection region for the tracking filters merely use the position measurement and will demonstrate the influence of amplitude information on tracking performance. Fig. \ref{fig_ground_truth}(b) shows the simulated SNR trajectories generated by the NCG distribution $f_d(d | d';\delta,\rho,c)$ with $\delta=1$, $\rho=0.999$ and $c=0.01$. Moreover, we employ a Gaussian distribution $p(d_{i+1}|d_i)=\mathcal{N}(d;d_i;4^2)$ as the proposal of the MH algorithm to generate $N_m=1000$ samples.\\
\indent The detection probability is fixed to $p_D=0.95$ for the GM-LMB filter in the absence of amplitude information, while it is calculated by \eqref{eqn_PD} for the HLMB filter. The detected target measurements are immersed in clutter which follows a Poisson RFS with an average intensity of $1.39\times 10^{-5}~\mathrm{m^{-2}}$ (i.e., 20 false alarms over the surveillance region per scan).  \\
\indent The birth model $\boldsymbol{\pi}_B=\{(r^\ell_B,p^\ell_B(x))\}_{\ell\in\mathbb{B}}$ can be obtained from measurements that do not assign to existing targets \cite{SBBK2014}. The birth distribution $p^\ell_B(x)$ can be decomposed into $p^\ell_B(\tilde{x} |d)$ and $p^\ell_B(d)$ using the Rao-Blackwellisation principle, where the conditional spatial density is in the form of GM $\sum_{i=1}^{J_B(\ell)} w_{B,i}^{\ell,\tilde{x}} \mathcal{N}(\tilde{x} ; m_{B,i}^{\ell}, P_{B, i}^{\ell})$ \cite{BDBB2012}, here we set $J_B(\ell)=5$. For SMC implementation of the target SNR, the initial particles are sampled uniformly from the assumed region $[10,40]$ dB with interval of 1 dB such that $N(\ell)=31$. Thus, $p^\ell_B(d)$ can be obtained by updating the initial particles with the amplitude measurement. For Gamma approximation of the target SNR, the prior SNR density $\gamma(d;\alpha_0,\beta_0)$ will be given by $\alpha_0\rightarrow1,\beta_0\rightarrow0$ for uniformative prior, or $\alpha_0=\beta_0=0$ for Jeffreys prior \cite{M2019}. If these inadequate priors are used to generate the initial sample of the MH algorithm, then the Markov chain $\mathcal{M}$ may not converge to the target density and lead to incorrect birth density $p^\ell_B(d)$. Instead, we use the updated particles from the aforementioned SMC initialization as the samples for the Gamma approximation of $p^\ell_B(d)$ given by \eqref{eqn_sample_mean_var} and \eqref{eqn_ab_update}.  \\
\indent In addition, the performance of the HLMB filter is measured using the optimal subpattern assignment (OSPA) metric modified with labeling error \cite{MUK2016,BBDB2011}, which takes into account the localization error, cardinality error and labeling error between the estimated states $\mathbf{Y}=\{(y_i,\ell^y_i)\}_{i=1}^m$ and the true states $\mathbf{X}=\{(x_j,\ell^x_j)\}_{j=1}^n$. Here $x$ and $y$ only represent the spatial states instead of the augmented states which also contain the target SNR. The accuracy of the estiamted SNR will be quantified independently by the root-mean-square error (RMSE). For $m\leq n$, the OSPA distance between these two state sets is given by
\begin{equation}\label{eqn_OSPA}
    \begin{aligned}
        \bar{d}_{p}^{c}(\mathbf{X}, \mathbf{Y})\triangleq & \left[\frac{1}{n}\left(\min _{\pi \in \Pi_n^m} \sum_{i=1}^{m} d_c\left(x_{i}, y_{\pi(i)}\right)^{p}\right.\right.\\
        &\left.\left.+\sum_{i=1}^m\left(d_\phi(\ell_i,\ell_{\pi^*(i)})\right)^p +c^{p}(n-m)\right)\right]^{\frac{1}{p}}\\
    \end{aligned}
\end{equation}
where $d_c(x_i,y_{\pi(i)})\triangleq\min(c,\vert\vert x_i-y_{\pi(i)}\vert\vert)$ is the cut-off distance with $c> 0$, $\vert\vert\cdot\vert\vert$ denotes the Euclidean distance, $\Pi_n^m$ represents the set of permutations of cardinality $m$ on $\{1,2,\dots,n\}$, and $p$ is the order. Besides, the labeling error $d_\phi(\ell_i,\ell_{\pi^*(i)})$ is defined as $d_\phi(a,b)\triangleq\phi\delta_a(b)$ with penalty $\phi>0$, and $\pi^*$ is the optimal arrangement which minimizes the summation of the cut-off distances in \eqref{eqn_OSPA}. For the case $m>n$, the OSPA distance is given by $\bar{d}_{p}^{c}(\mathbf{X}, \mathbf{Y})=\bar{d}_{p}^{c}(\mathbf{Y}, \mathbf{X})$. Here we set $c=30$ m, $\phi=30$ m and $p=1$. 
\subsection{Results}
\begin{figure*}[t]
    \centering
    \subfloat[GM-LMB]{
        \includegraphics[width=1.4in]{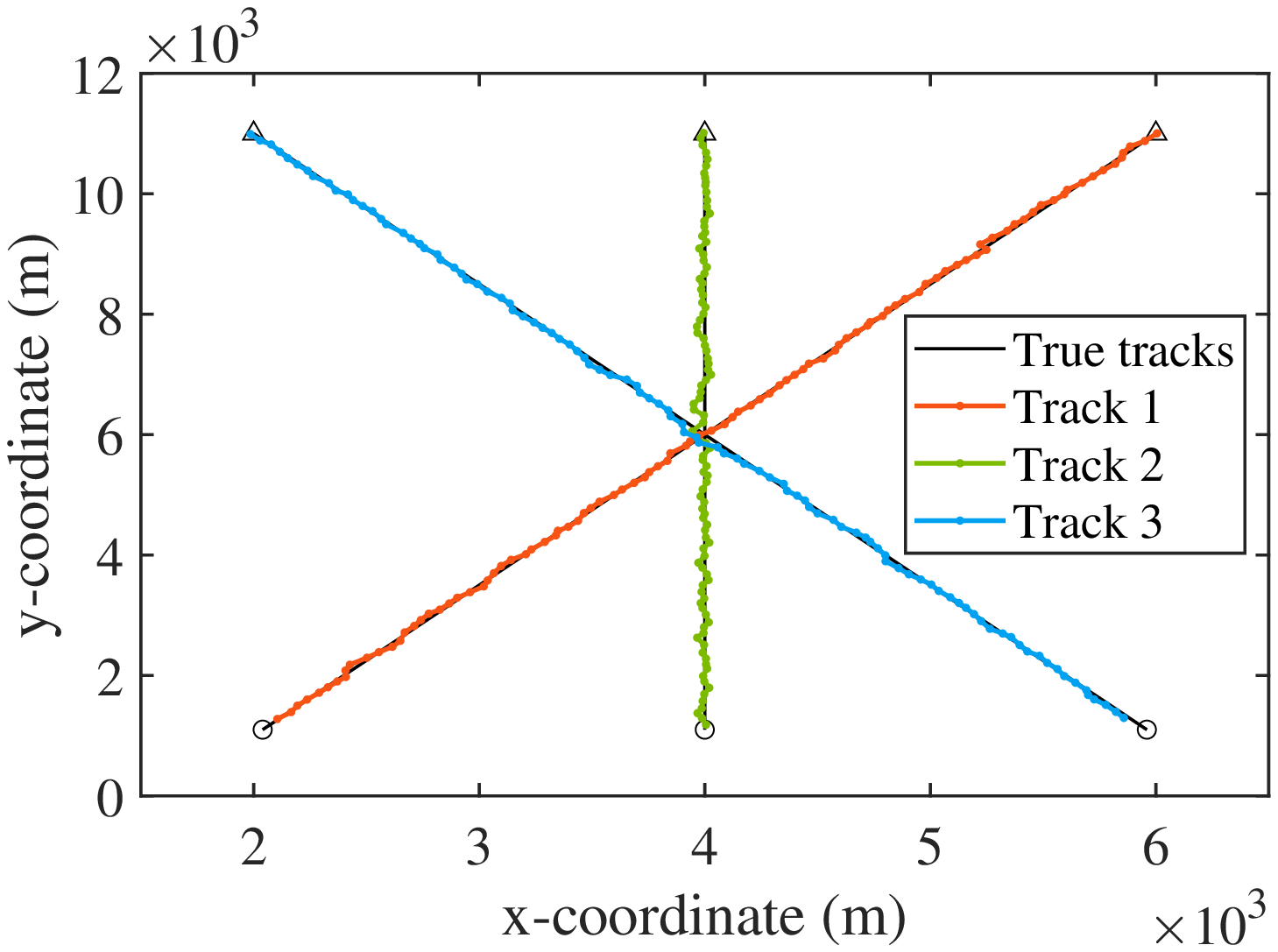}
    }\hspace{-3.5mm}
    \subfloat[GM-LMB-K]{
        \includegraphics[width=1.4in]{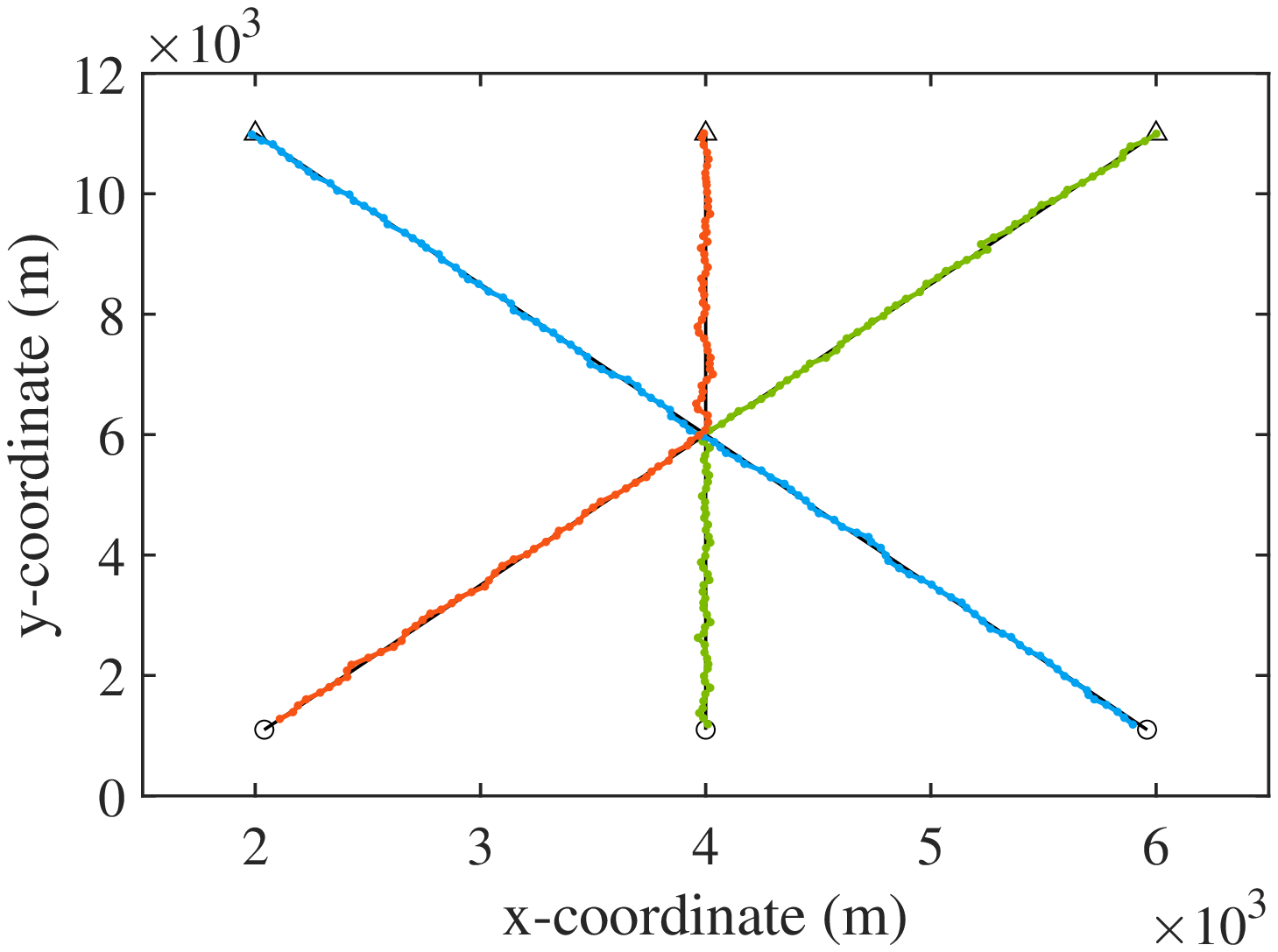}
    }\hspace{-3.5mm}
    \subfloat[GM-LMB-M]{
        \includegraphics[width=1.4in]{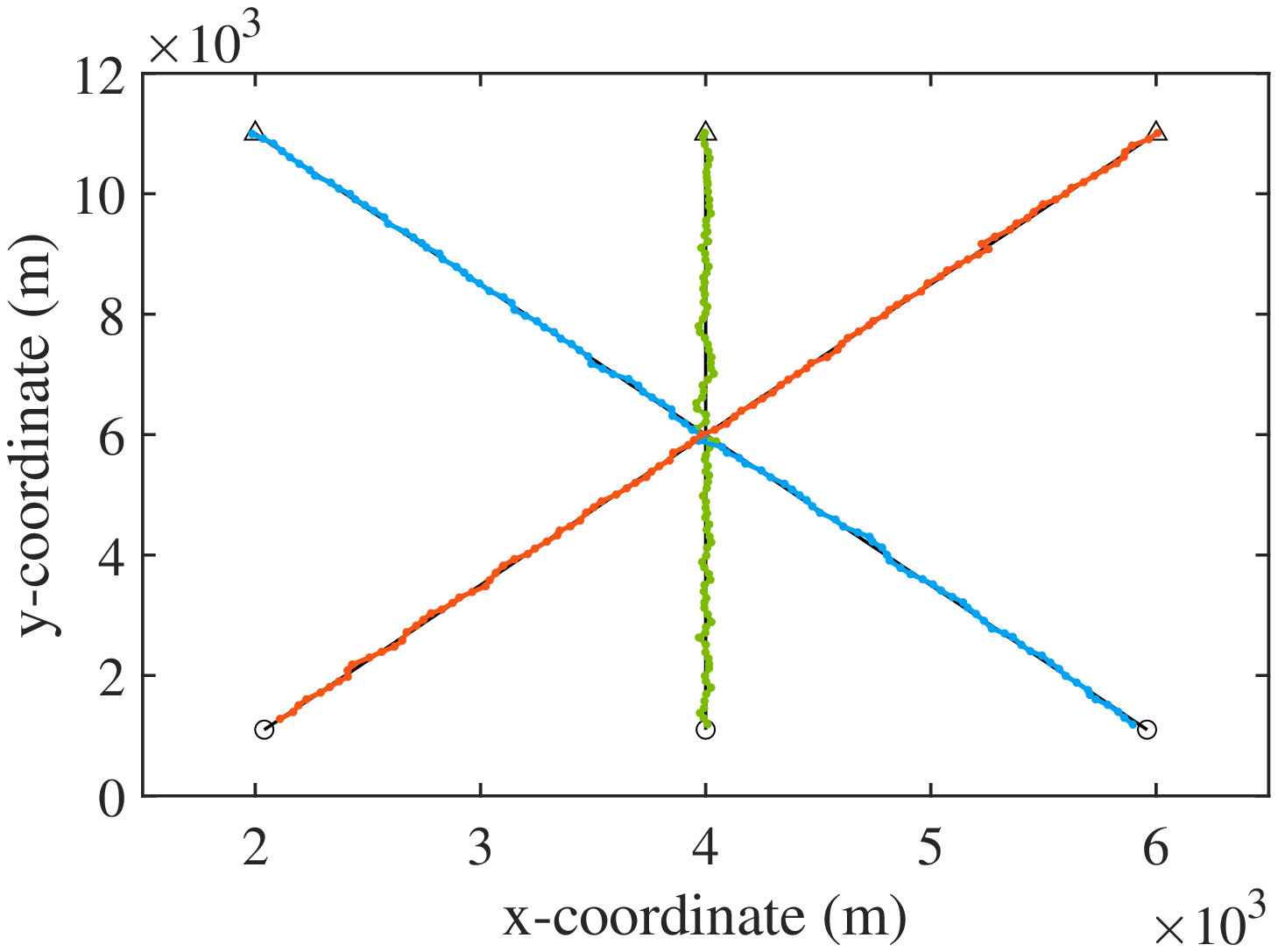}
    }\hspace{-3.5mm}
    \subfloat[GM/SMC-HLMB]{
        \includegraphics[width=1.4in]{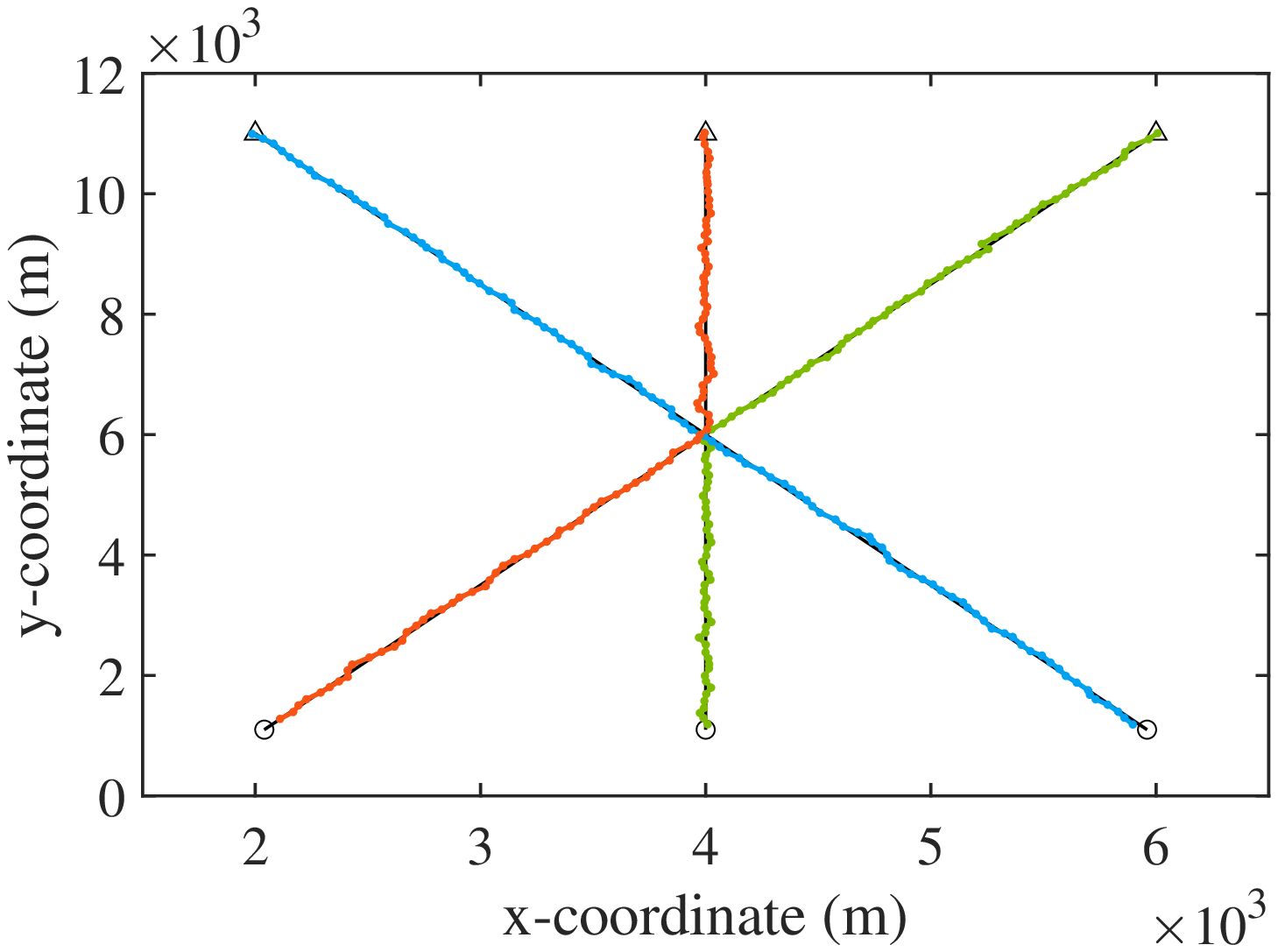}
    }\hspace{-3.5mm}
    \subfloat[GM/G-HLMB]{
        \includegraphics[width=1.4in]{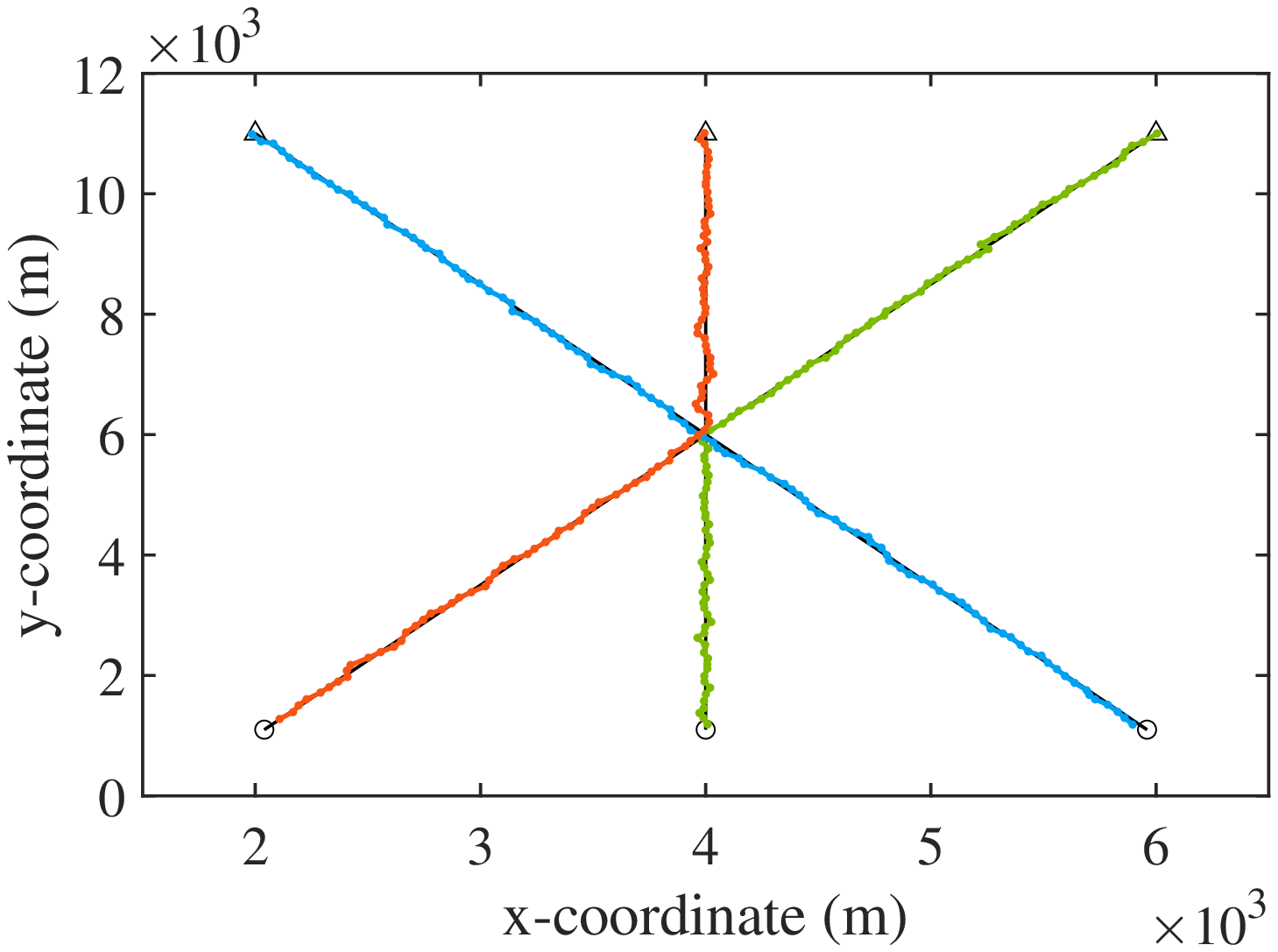}
    }
    \caption{The estimated tracks of the testing algorithms with Swerling 1 amplitude likelihood from a single run.}
    \label{fig_est_track}
\end{figure*}
\begin{figure}[h]
    \centering
    \subfloat[Swerling 1]{
        \includegraphics[width=1.65in]{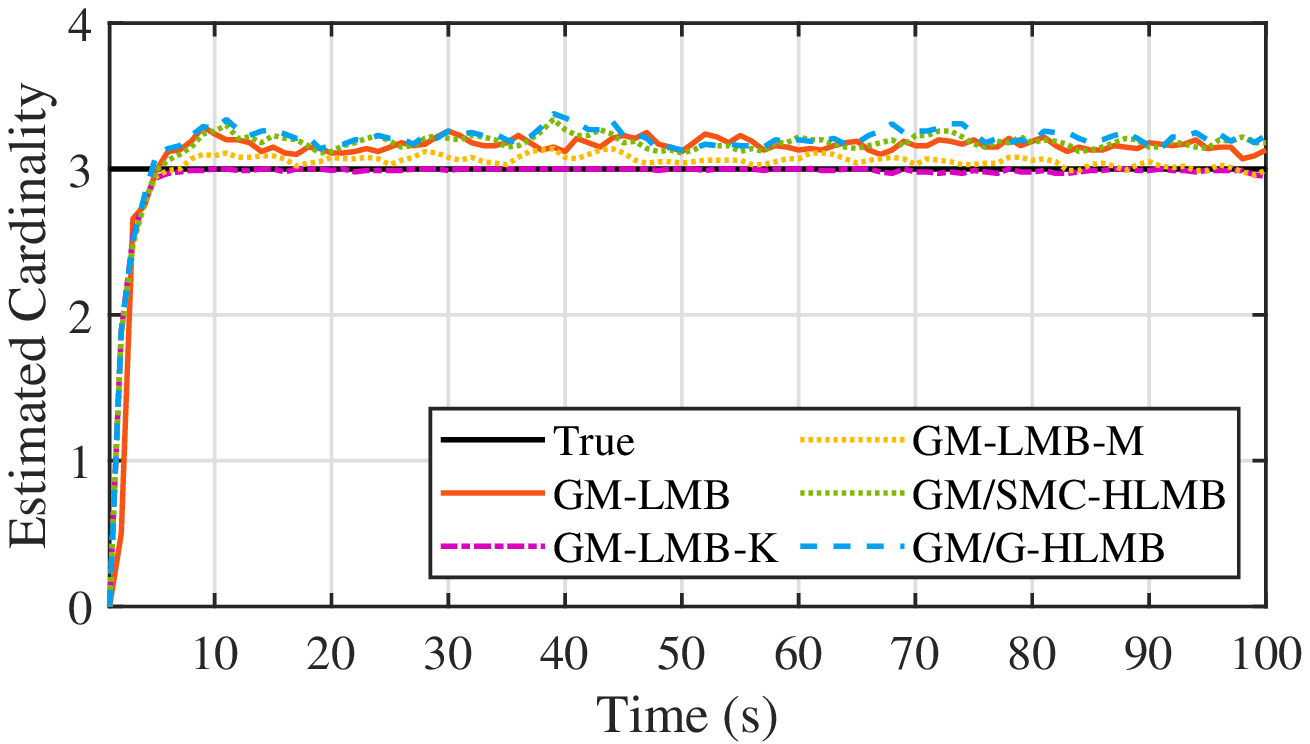}
    }\hspace{-3mm}
    \subfloat[Swerling 3]{
        \includegraphics[width=1.65in]{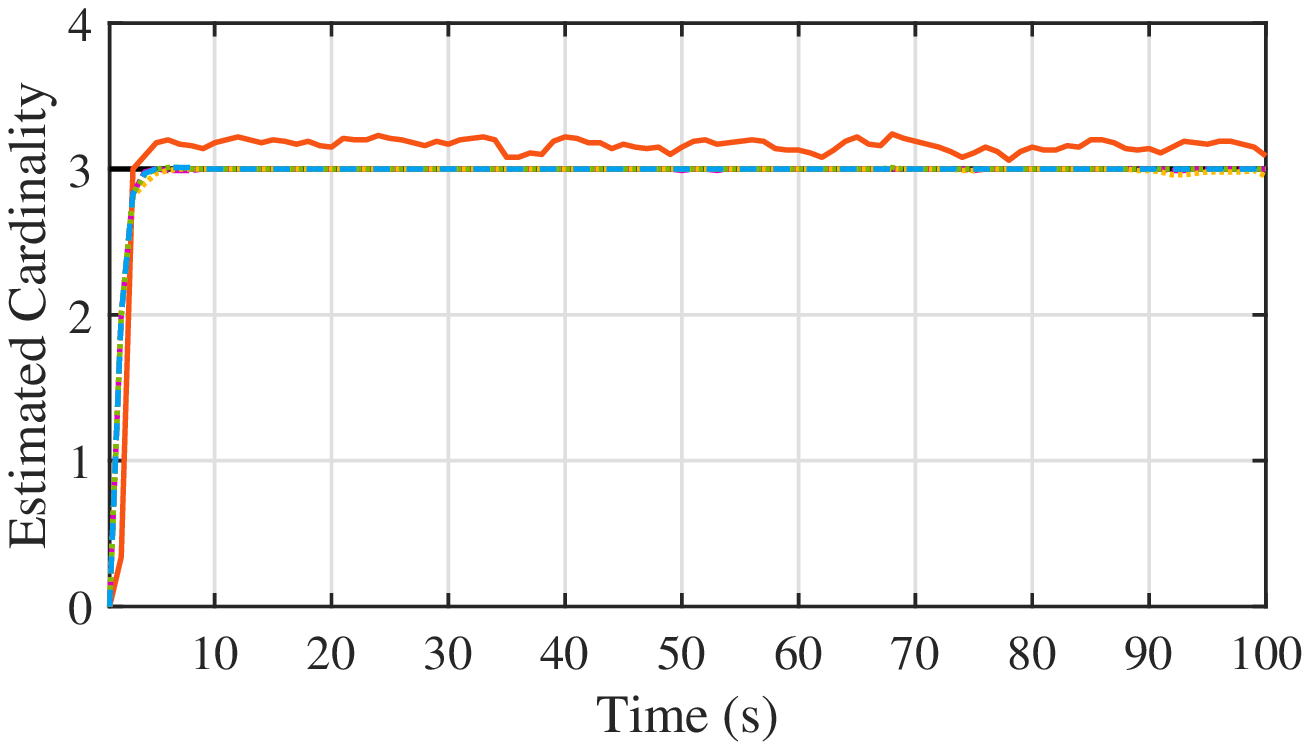}
    }
    \caption{Cardinality estimates with Swerling 1 and 3 amplitude likelihoods.}
    \label{fig_card_est}
\end{figure}

\indent Fig. \ref{fig_est_track} displays the estimated tracks of the testing algorithms produced by a single run with Swerling 1 amplitude likelihood. As expected, the GM-LMB filter without amplitude exhibits obvious track switchings after the intersection. Indeed, the latter half of track 1 in Fig. \ref{fig_est_track}(a) represents target 2 while the latter half of track 2 belongs to target 1. The GM-LMB-M filter also suffers from track switchings since the marginalized likelihoods of one amplitude measurement to different targets are the same, leading to no discrimination between targets. However, algorithms with amplitude information, based on known or estimated SNR, successively track each target and avoid track switchings. \\
\begin{figure}[t]
    \centering
    \subfloat[Overall OSPA error]{
        \includegraphics[width=1.65in]{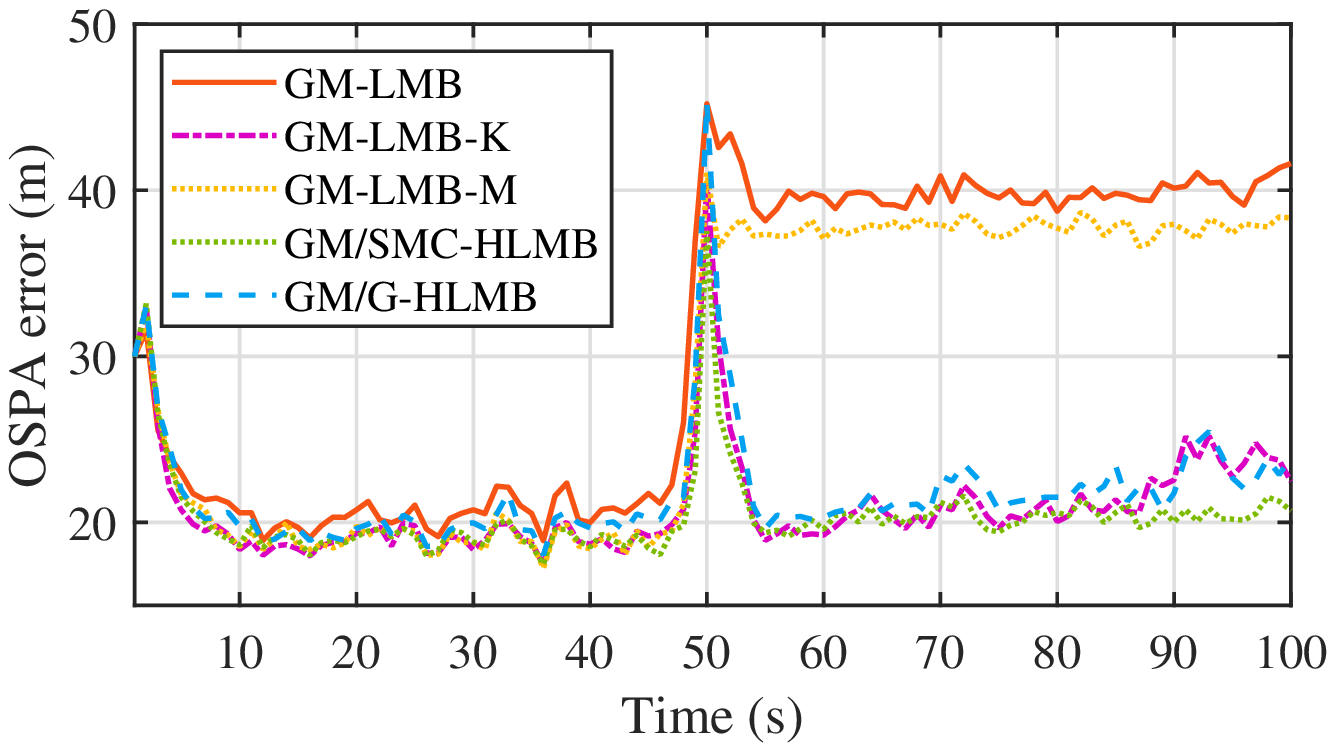}
    }\hspace{-3mm}
    \subfloat[Localization error]{
        \includegraphics[width=1.65in]{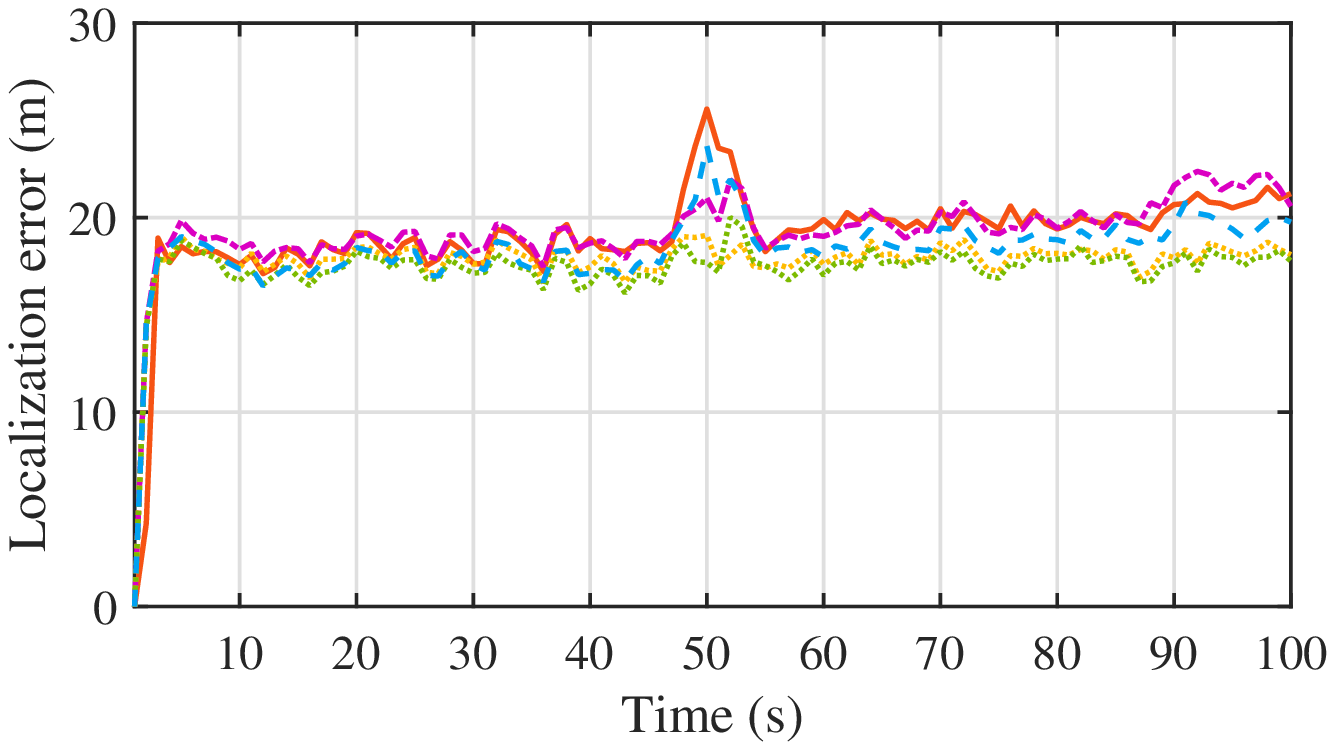}
    }\vspace{-0mm}
    \subfloat[Labeling error]{
        \includegraphics[width=1.65in]{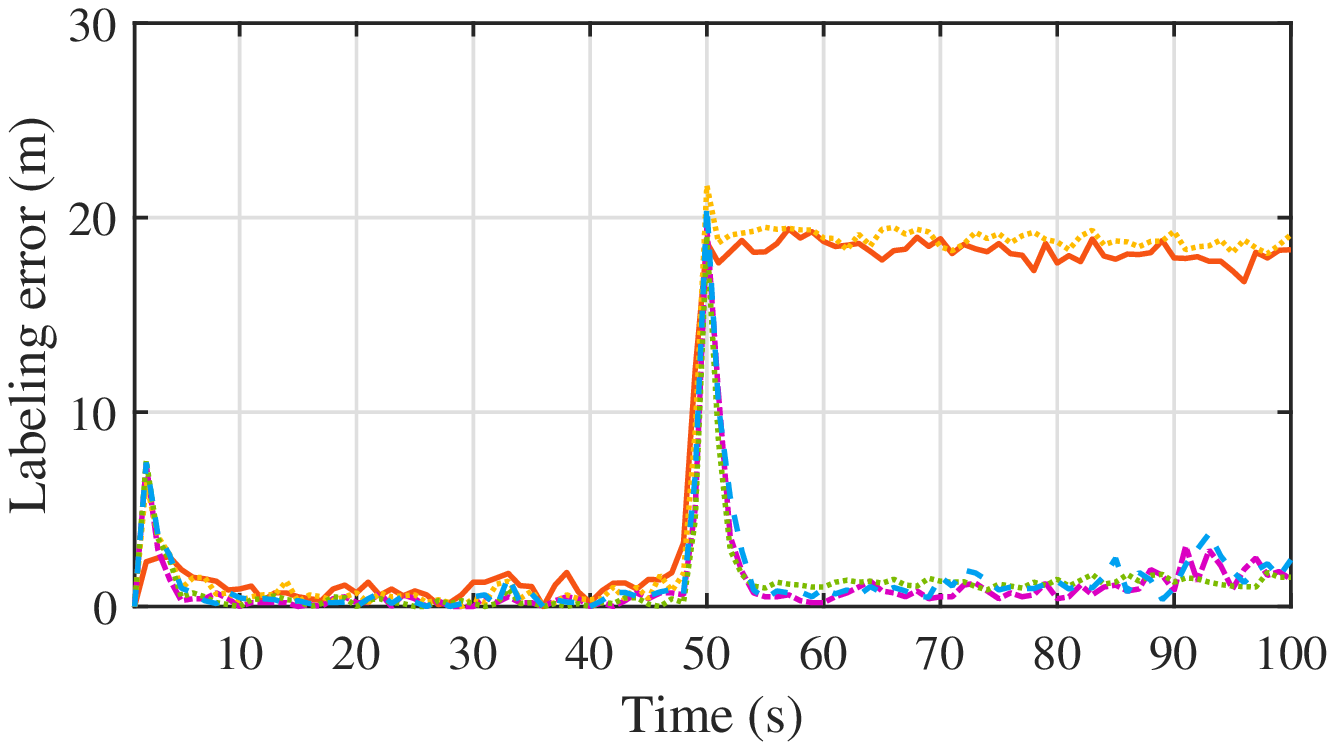}
    }\hspace{-3mm}
    \subfloat[Cardinality error]{
        \includegraphics[width=1.65in]{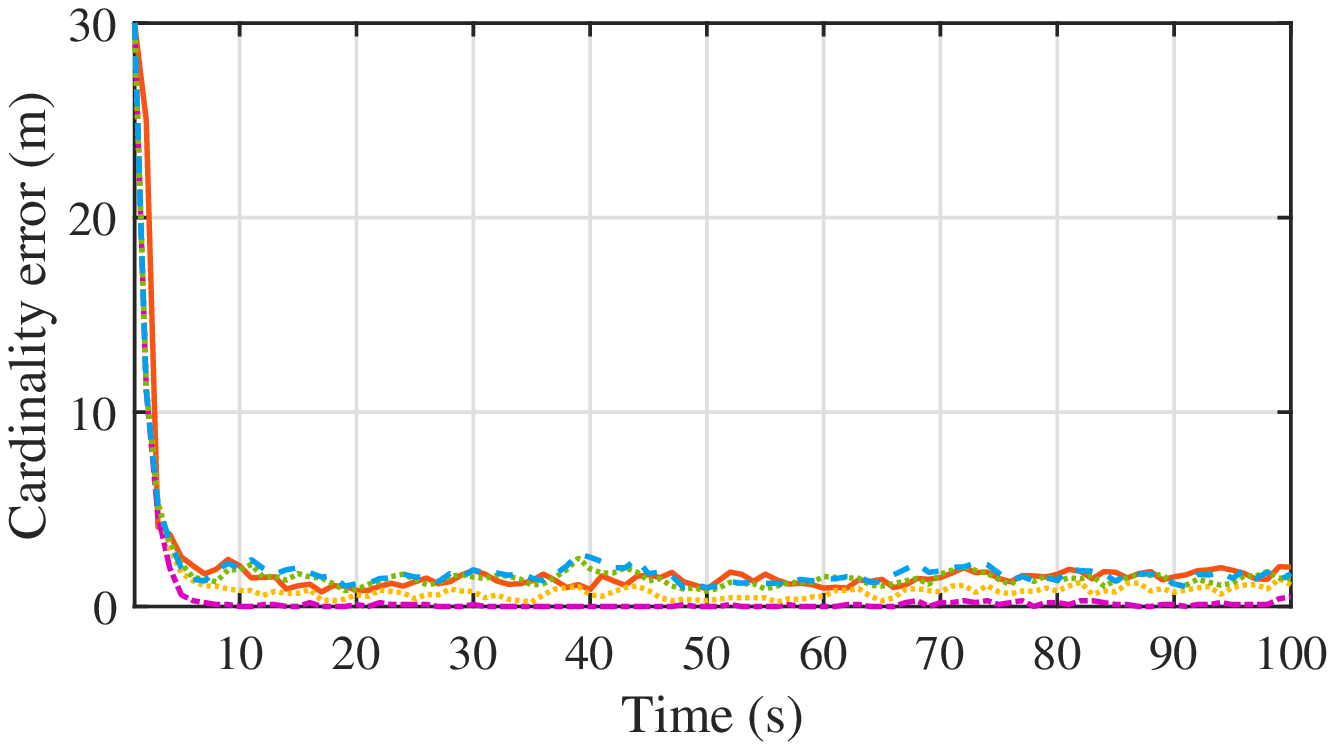}
    }
    \caption{Average OSPA errors and the contributing components of the testing algorithms with Swerling 1 amplitude likelihood.}
    \label{fig_ospal_s1}
\end{figure}
\begin{figure}[t]
    \centering
    \subfloat[Overall OSPA error]{
        \includegraphics[width=1.65in]{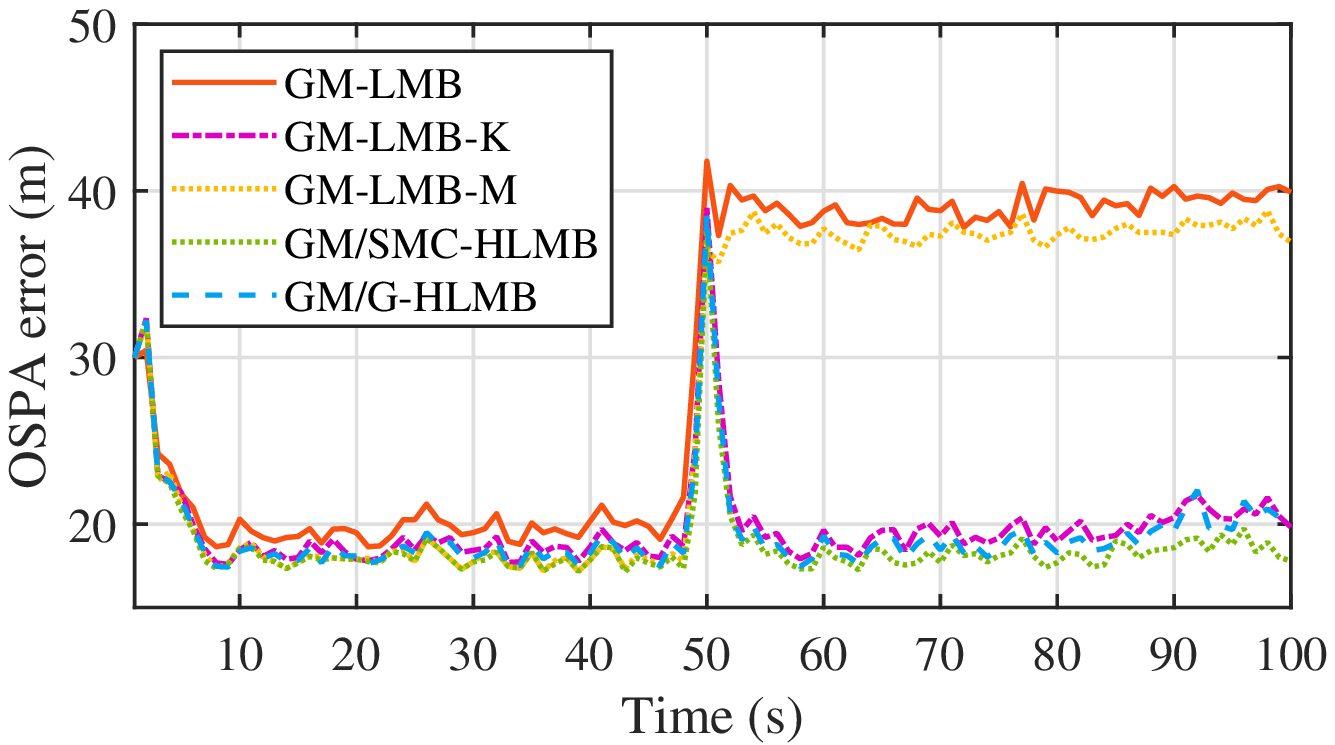}
    }\hspace{-3mm}
    \subfloat[Localization error]{
        \includegraphics[width=1.65in]{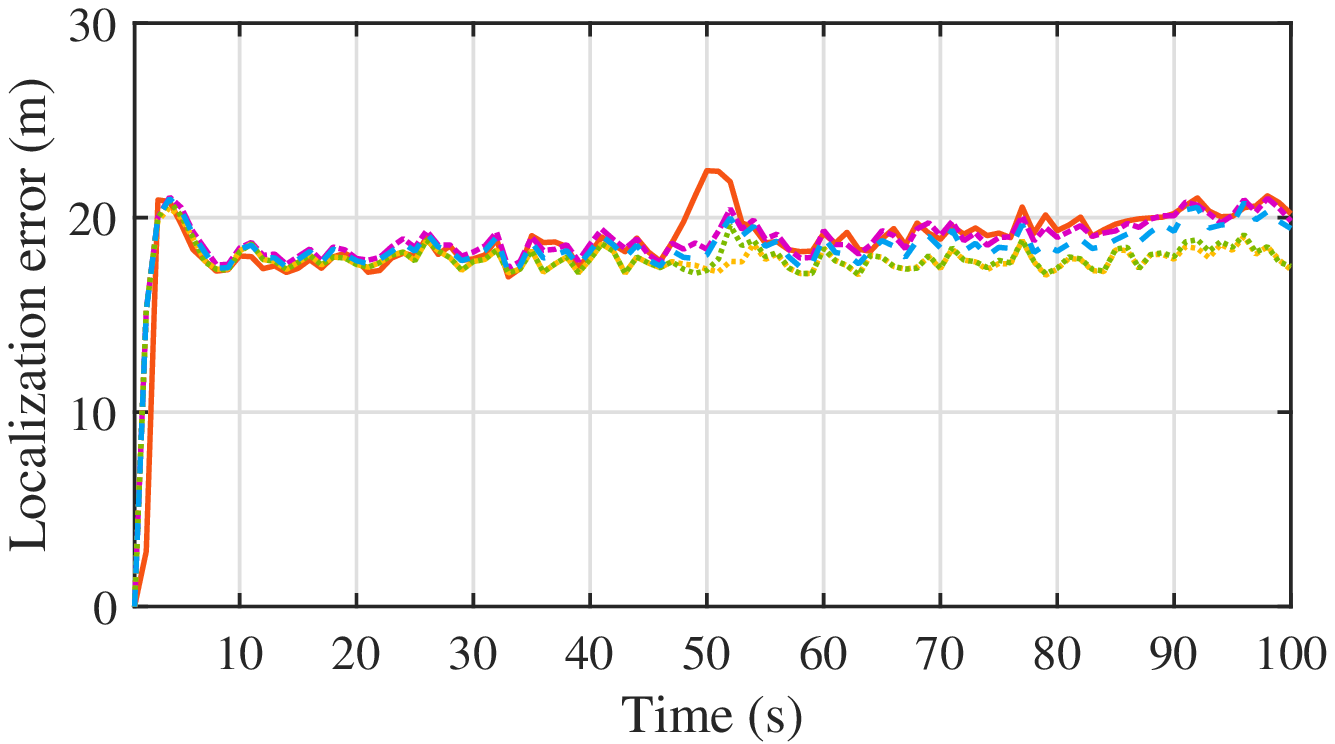}
    }\vspace{0mm}
    \subfloat[Labeling error]{
        \includegraphics[width=1.65in]{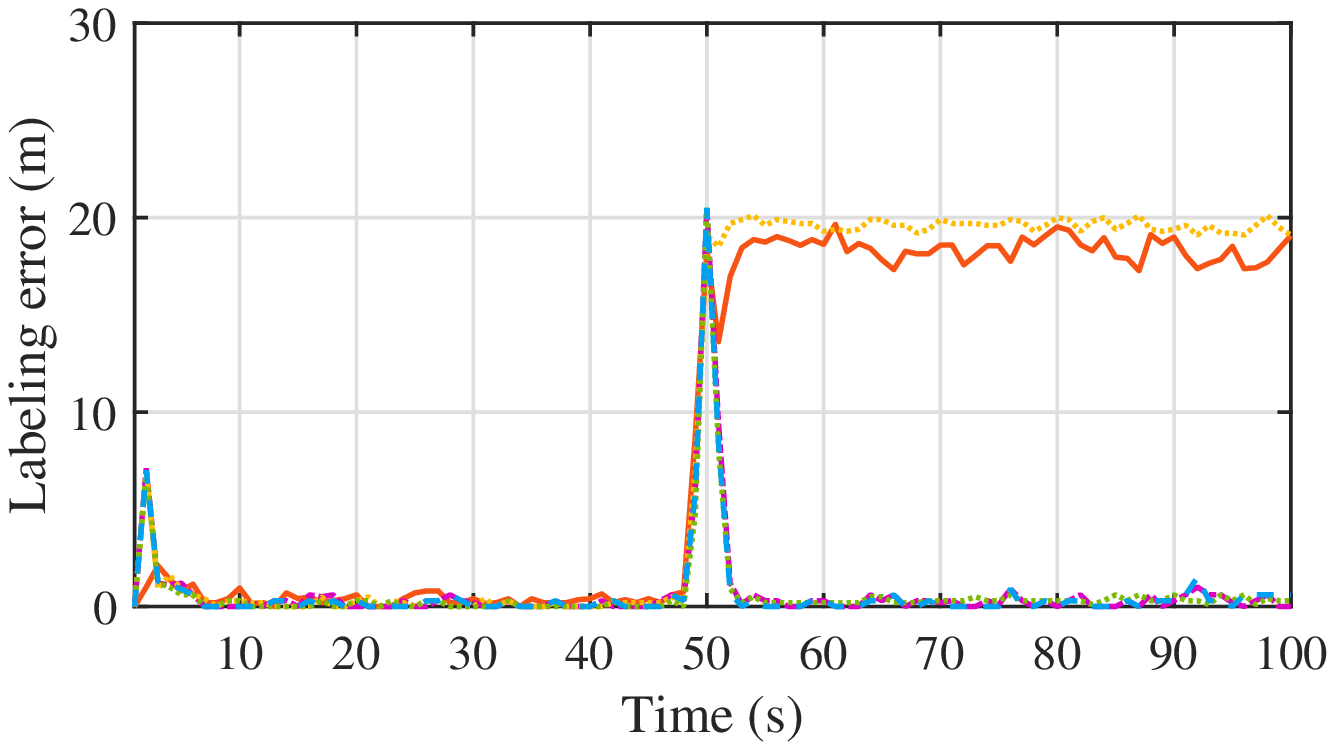}
    }\hspace{-3mm}
    \subfloat[Cardinality error]{
        \includegraphics[width=1.65in]{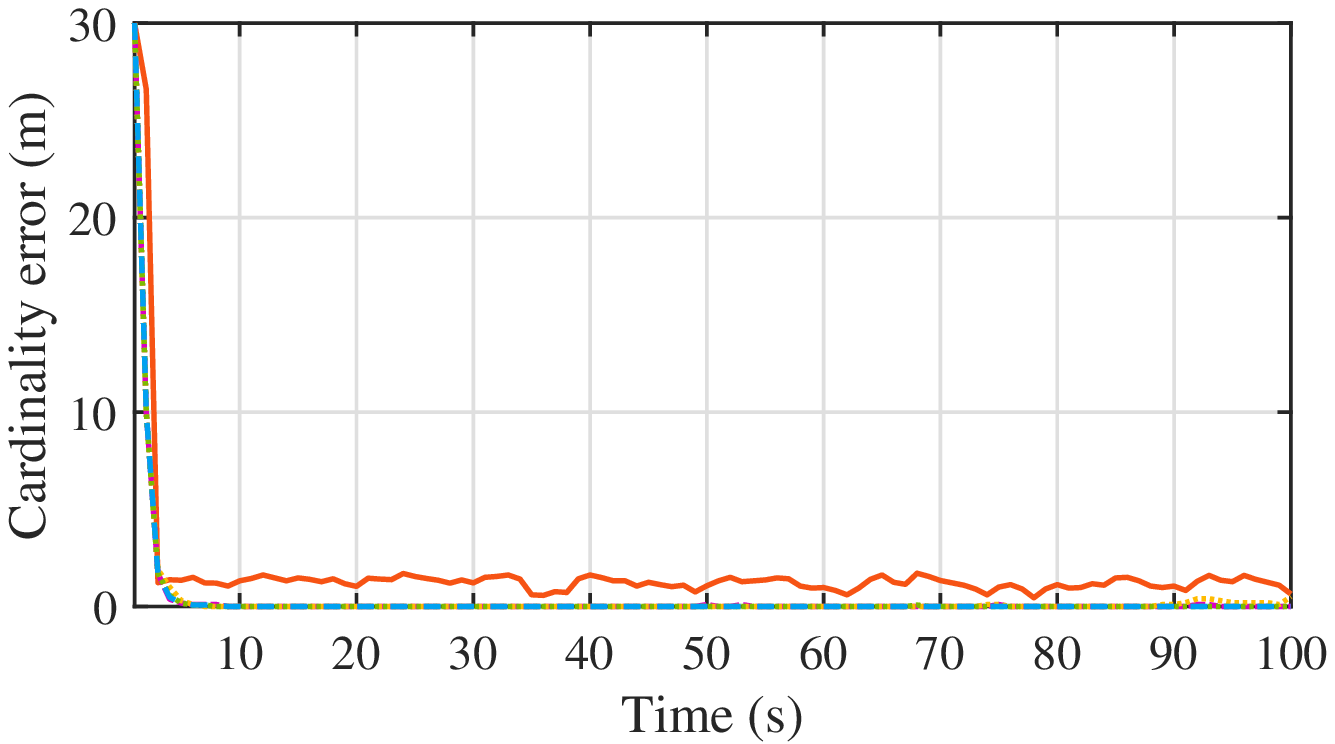}
    }
    \caption{Average OSPA errors and the contributing components of the testing algorithms with Swerling 3 amplitude likelihood.}
    \label{fig_ospal_s3}
\end{figure}
\indent Figs. \ref{fig_ospal_s1} and \ref{fig_ospal_s3} shows the OSPA distances of the testing algorithms, averaged over 100 Monte Carlo runs, with Swerling 1 and 3 amplitude likelihoods. Also shown are the contributing localization, labeling and cardinality components. Before the crossing, all algorithms exhibit stable performance and the overall OSPA errors of algorithms with amplitude are smaller than that of GM-LMB. This improvement results from that amplitude helps filter out clutters and distinguish targets from the remaining clutters. As the targets enter the intersection region, the overall OSPA errors of all algorithms increase dramatically due to the surge of the labeling errors, which results from the uncertainty of measurement to track association. After the intersection, the labeling errors of the GM-LMB and GM-LMB-M remains high because of the track switchings shown in Fig. \ref{fig_est_track}(a) and Fig. \ref{fig_est_track}(c), while the labeling errors of other methods decrease after a few time steps. This difference indicates that algorithms using kinematic measurement only or imprecise amplitude likelihood can hardly recognize the changed directions, and algorithms with precise amplitude likelihood can sense and adapt to the changed directions. \\
\indent It can also be observed from Figs. \ref{fig_ospal_s1} and \ref{fig_ospal_s3} that the labeling errors of algorithms with Swerling 3 amplitude likelihood are smaller than with Swerling 1 after the intersection. This is because that Swerling 3 amplitude likelihood provides stronger discriminations between targets to mitigate track switchings, see Fig. \ref{fig_amplitude_lh}. In terms of the localization error, the GM/SMC-HLMB fiter outperforms other methods since it employs more Gaussian components. The cardinality errors are all negligible after a few scans. Besides, the overall performance of the GM/SMC-HLMB and GM/G-HLMB filters is comparable to that of the GM-LMB-K filter, indicating the efficacy of the SNR estimators. \\
\begin{figure}[t]
    \centering
    \subfloat[GM/SMC-HLMB Swerling 1]{
        \includegraphics[width=1.6in]{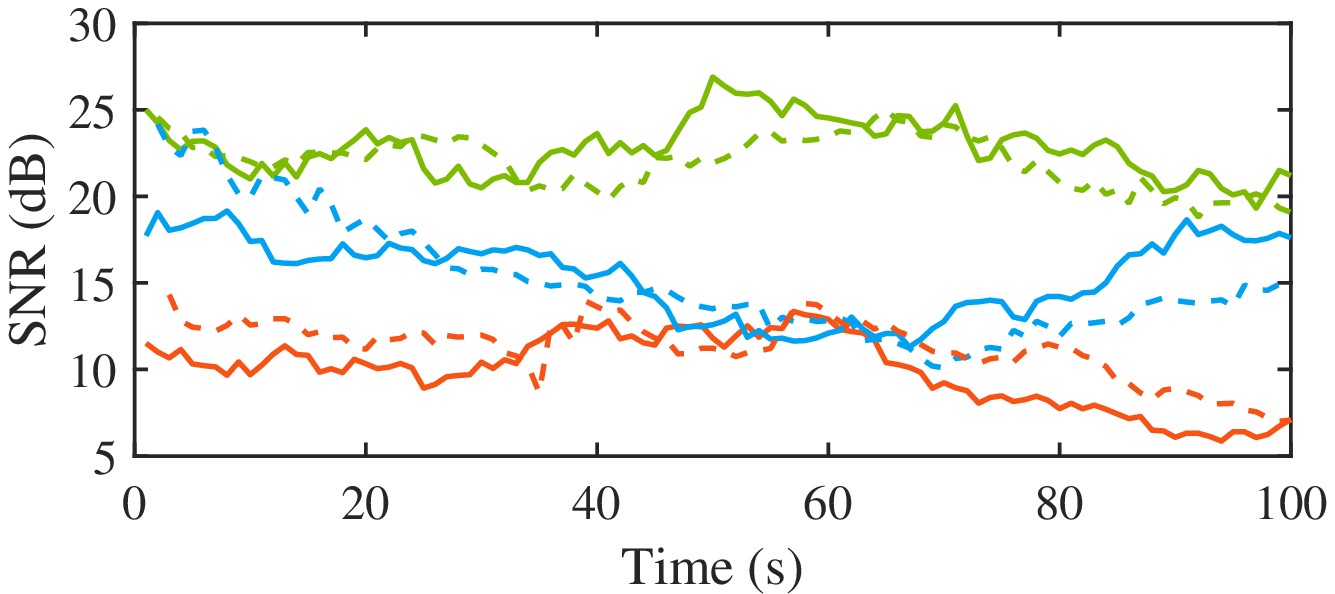}
    }
    \subfloat[GM/G-HLMB Swerling 1]{
        \includegraphics[width=1.6in]{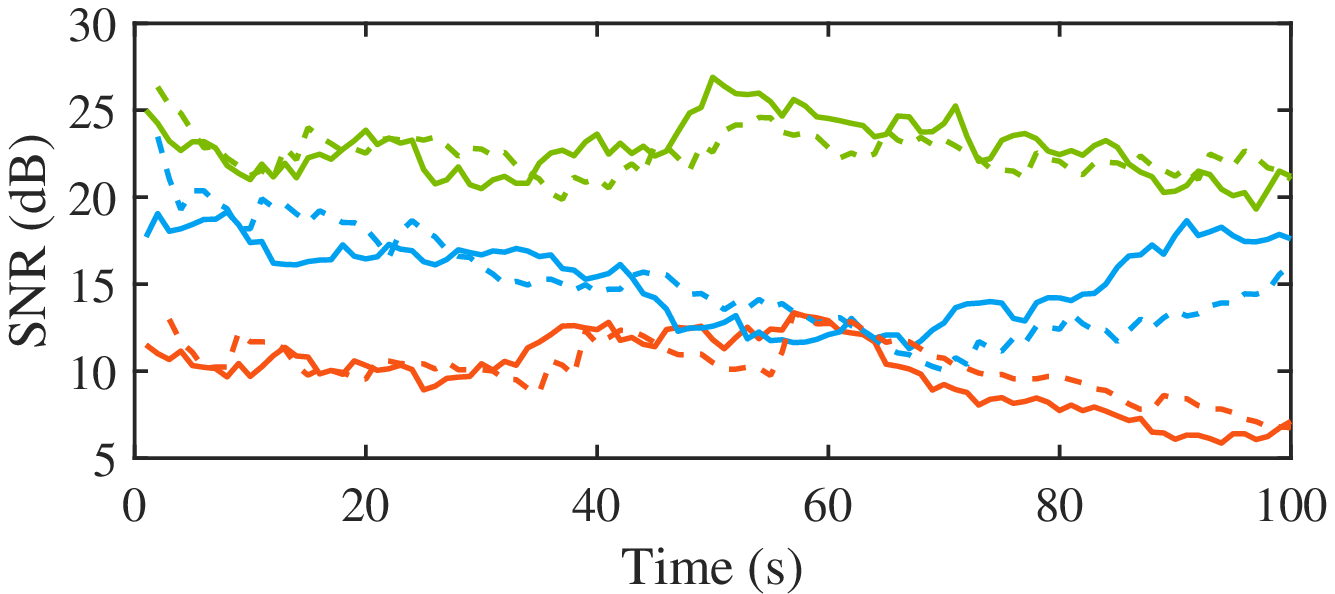}
    }
    \vspace{-3mm}
    \subfloat[GM/SMC-HLMB Swerling 3]{
        \includegraphics[width=1.6in]{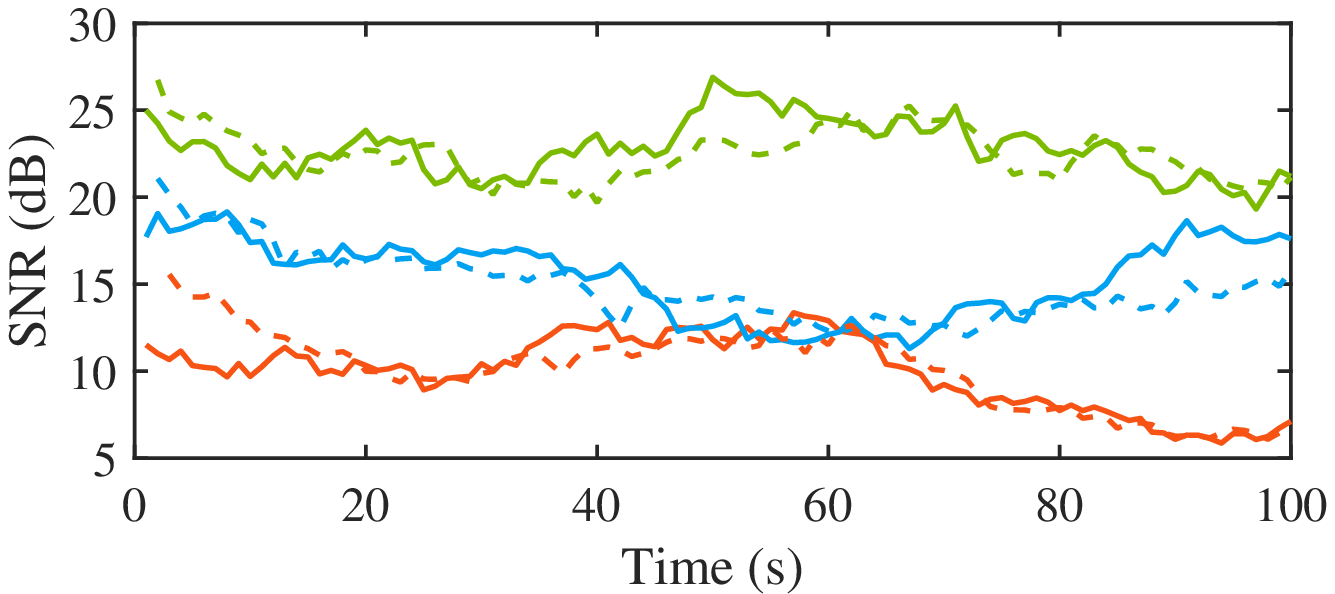}
    }
    \subfloat[GM/G-HLMB Swerling 3]{
        \includegraphics[width=1.6in]{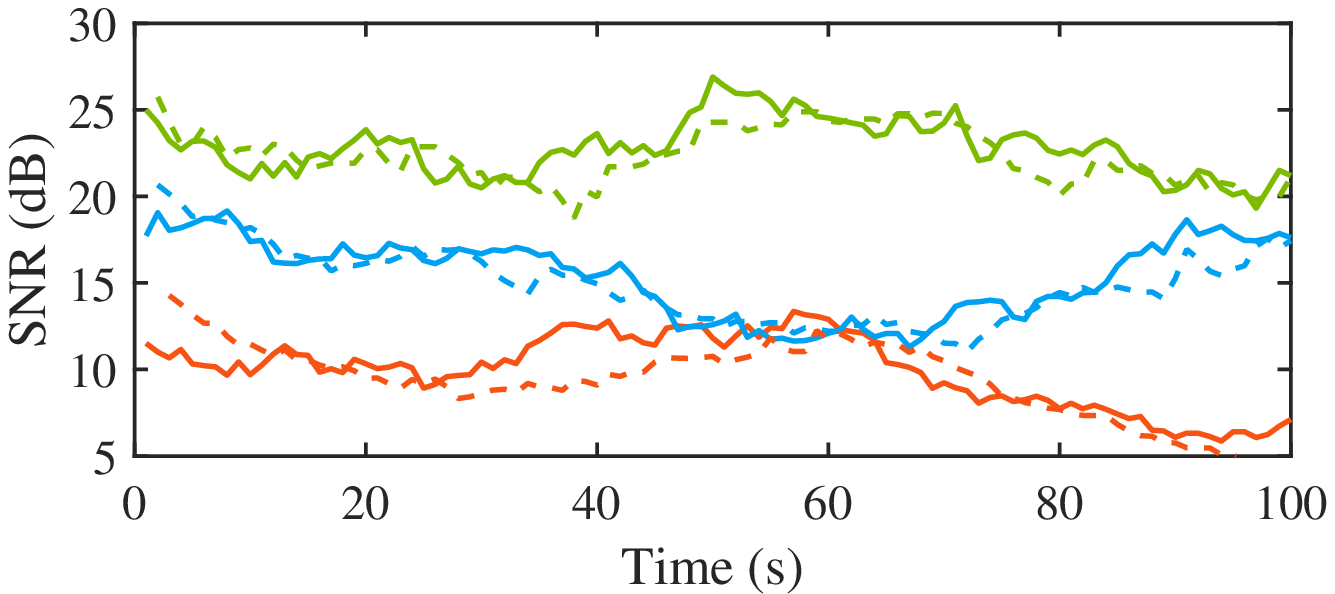}
    }
    \caption{The estimated SNR trajectories from a single run with Swerling 1 and 3 amplitude likelihoods. (Solid lines represent the true trajectories and dashed lines represent the estimated.)}
    \label{fig_snr_est}
\end{figure}
\begin{figure}[t]
    \centering
    \includegraphics[width=3in]{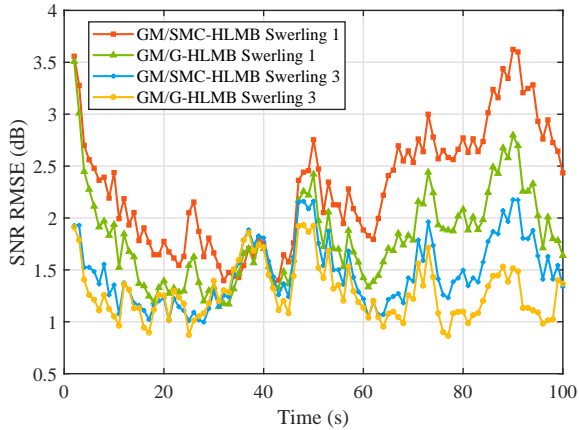}
    \caption{The SNR RMSEs of the GM/SMC-HLMB and GM/G-HLMB filters with Swerling 1 and 3 amplitude likelihoods.}
    \label{fig_snr_rmse}
\end{figure}
\indent To analyze the accuracy of the estimated SNR obtained from GM/SMC-HLMB and GM/G-HLMB, Fig. \ref{fig_snr_est} displays an exemplary SNR trajectories and Fig. \ref{fig_snr_rmse} shows the RMSE of the SNR averaged over 100 Monte Carlo trials. It can be seen that GM/G-HLMB provides more accurate estimation of the target SNR than GM/SMC-HLMB for either amplitude likelihood. This is owing to that GM/G-HLMB employs more smaples to approximate the posterior SNR density. These two algorithms with Swerling 3 amplitude likelihood outperforms with Swerling 1 since that Swerling 1 target exhibits severer amplitude fluctuation than Swerling 3 target, see Fig. \ref{fig_amplitude_lh}, which hinders the estimation of the target SNR. The OSPA errors and the SNR RMSEs, shown in Figs. \ref{fig_ospal_s1}, \ref{fig_ospal_s3} and \ref{fig_snr_rmse}, are averaged over time in Tables \ref{tab_ospa_s1} and \ref{tab_ospa_s3} for reference. \\
\begin{table}[tb]
    \renewcommand{\arraystretch}{1.1}
    \centering
    \caption{The estimation errors with Swerling 1 amplitude likelihood}
    \label{tab_ospa_s1}
    \resizebox{\linewidth}{!}
    {
        \begin{tabular}{cccccc} \hline\hline
            Algorithm & \makecell[c]{OSPA \\(m)} & \makecell[c]{Localization\\(m)} & \makecell[c]{Labeling\\(m)} & \makecell[c]{Cardinality\\(m)} &\makecell[c]{SNR \\(dB)} \\ \hline
            GM-LMB & 31.07 & 19.09 & 9.95 & 2.02 & - \\ 
            GM-LMB-K & 20.96 & 19.27 & 1.13 & 0.57 & - \\
            GM-LMB-M & 29.13 & 17.76 & 10.16 & 1.21 & - \\
            GM/SMC-HLMB & 20.44 & 17.38 & 1.19 & 1.87 & 2.27\\ 
            GM/G-HLMB & 21.78 & 18.35 & 1.37 & 2.06 & 1.79 \\ \hline\hline
        \end{tabular}
    }
\end{table}
\begin{table}[tb]
    \renewcommand{\arraystretch}{1.1}
    \centering
    \caption{The estimation errors with Swerling 3 amplitude likelihood}
    \label{tab_ospa_s3}
    \resizebox{\linewidth}{!}
    {
        \begin{tabular}{cccccc} \hline\hline
            Algorithm & \makecell[c]{OSPA \\(m)} & \makecell[c]{Localization\\(m)} & \makecell[c]{Labeling\\(m)} & \makecell[c]{Cardinality\\(m)} &\makecell[c]{SNR \\(dB)} \\ \hline
            GM-LMB & 30.04 & 18.66 & 9.62 & 1.76 & - \\ 
            GM-LMB-K & 19.84 & 18.74 & 0.67 & 0.43 & - \\
            GM-LMB-M & 28.38 & 17.69 & 10.23 & 0.46 & - \\
            GM/SMC-HLMB & 18.81 & 17.75 &  0.63 & 0.43 & 1.46 \\ 
            GM/G-HLMB & 19.44 & 18.38 & 0.63 & 0.43 & 1.27 \\ \hline\hline
        \end{tabular}
    }
\end{table}
\indent A comparison of the average computational time of the testing algorithms with MATLAB implementation on a Core I9-12900K CPU is presented in Table \ref{tab_computational_time}. The average execution time of the GM-LMB is indeed 14.94 s and 15.31 s for Swerling 1 and 3 amplitude likelihoods. The GM-LMB-M filter appears to be the most computationally efficient method with amplitude due to that the marginalized likelihood involves no estimating of the target SNR. It is even faster than GM-LMB for Swerling 1 target since it estimates the cardinality more accurately, see the cardinality estimates in Fig. \ref{fig_card_est}(a). The computational improvement of GM/G-HLMB over GM/SMC-HLMB is owing to that the kinematic state is calculated based on the estimated SNR rather than each particle. Algorithms with Swerling 1 are slower than with Swerling 3 since they overestimate the cardinality, as shown in Fig. \ref{fig_card_est}. However, the GM-LMB-M filter is slower for Swerling 3 target since the corresponding marginalized likelihood involves numerical integral, whereas the marginalized likelihood for Swerling 1 target is analytic.
\begin{table}[h]
    \renewcommand{\arraystretch}{1.1}
    \setlength\tabcolsep{3pt}
    \centering
    \caption{Computational Time relative to the GM-LMB}
    \label{tab_computational_time}
    \resizebox{\linewidth}{!}
    {
        \begin{tabular}{cccccc} \hline\hline
            Model & GM-LMB & GM-LMB-K & GM-LMB-M & GM/SMC-HLMB & GM/G-HLMB \\ \hline
            Swerling 1 & 1 & 1.20 & 0.90 & 3.17 & 1.59 \\ 
            Swerling 3 & 1 & 1.19 & 1.14 & 2.47 & 1.39 \\ \hline\hline
        \end{tabular}
    }
\end{table}
\vspace{-0.3cm}
\section{Conclusion}\label{sec_six}
\indent In this paper we introduce a HLMB filter with amplitude information for tracking targets with unknown and fluctuating SNR. The fluctuation of the target SNR is modeled by an ARG process while Rayleigh and one-dominant-plus-Rayleigh amplitude likelihoods are considered for Swerling 1 and 3 targets, respectively. The proposed HLMB filter utilizes the Rao-Blackwellisation principle to decompose the augmented state density into two parts: the densities of the target SNR and the kinematic state conditioned on SNR. An approximate Gamma recursion for estimating the target SNR is developed based on Laplace transform and MCMC method, and a GM filter is employed for the conditional spatial state. Simulation results obtained from a scenario containing crossing targets show that the GM/SMC-HLMB and GM/G-HLMB filters mitigate the track switching problem compared with the GM-LMB and GM-LMB-M filters and provide comparable performance to the GM-LMB-K filter for either Swerling 1 or 3 target. The GM/SMC-HLMB filter outperforms GM/G-HLMB in terms of localization error, while GM/G-HLMB exceeding in labeling and SNR accuracy and computational efficiency. Future work might concentrate on adaptive amplitude model such that the hybrid filter can simultaneously track Swerling 1 and 3 targets.
\bibliographystyle{IEEEtran}
\bibliography{reference.bib}
\end{document}